UATP /08-03

# Compressible or incompressible blend of interacting monodisperse star and linear polymers near a surface

*Richard Batman*[†] *and P. D. Gujrati*

Department of Physics, The University of Akron, Akron, OH 44325

ABSTRACT

We consider a lattice model of a mixture of repulsive, attractive, or neutral monodisperse star (species A) and linear (species B) polymers with a third monomeric species C, which may represent free volume. The mixture is next to a hard, infinite plate whose interactions with A and C can be attractive, repulsive, or neutral. These two interactions are the only parameters necessary to specify the effect of the surface on all three components. We numerically study monomer density profiles using the method of Gujrati and Chhajer that has already been previously applied to study polydisperse and monodisperse linear-linear blends next to surfaces. The resulting density profiles always show an enrichment of linear polymers in the immediate vicinity of the surface, due to entropic repulsion of the star core. However, the integrated surface excess of star monomers is sometimes positive, indicating an overall enrichment of stars. This excess increases with the number of star arms only up to a certain critical number and decreases thereafter. The critical arm number increases with compressibility (bulk concentration of C). The method of Gujrati and Chhajer is computationally ultrafast and can be carried out on a PC, even in the incompressible case, when simulations are unfeasible. Calculations of density profiles usually take less than 20 minutes on PCs.

[†] Current address: Louisiana School for Math, Science, and the Arts, 715 University Parkway, Natchitoches, Louisiana 71457.

## INTRODUCTION

There has been much recent interest in the effect of surfaces and interfaces on the proximate composition of polymer mixtures, due to the importance of surface properties to technologies such as adhesion, lubrication, and biocompatibility. The ultimate aim of research in this field is to develop a theory that will allow one to calculate in advance the profiles of composition and various other densities as a function of depth below the surface. The usual approach to such a calculation involves mean-field (MF) approximations to calculate the partition function (PF) of a mixture modeled on a cubic lattice. A fundamentally different approach has been recently developed by Gujrati and Chhajer[1-6] at the University of Akron, in which the cubic lattice is replaced by a recursive structure on which the PF is calculated exactly, without resorting to the Random Mixing Approximation (RMA). The advantage of this approach over MF theory is that it captures more local correlation between monomers, allowing one to produce profile features that cannot be generated by a MF approach, but are observed in simulations. It also produces results in much less time than is required for a Monte Carlo (MC) simulation, and in cases that cannot be feasibly handled by them, such as fully-packed lattices and free energy or entropy calculations. Most of the results presented in this paper required less than 20 minutes to produce.

The aim of this paper is to apply the Gujrati-Chhajer approach to compressible blends of star and linear polymers with fixed numbers of arms and segments confined between two hard parallel plates. The separation between them is adjusted to be large enough that the mixture reaches its bulk composition midway between the plates. Therefore the system is equivalent to a semi-infinite system next to a single surface. We will show the effect on surface segregation of the architectural difference between the components, determined by the number of arms in the star polymer, and study how the effect of architecture on surface segregation depends on the density of free volume (percentage of void sites) in the uniform bulk composition.

The reader should be warned of some limitations of the method. Although its avoidance of the RMA makes it more reliable than the MF approach, the Bethe lattice that replaces the original cubic lattice in this work possesses much weaker correlation than the original lattice and cannot generate non-classical critical exponents near critical points. The theory generated by the Gujrati-Chhajer approach applied to the Bethe lattice must therefore, in this latter sense, be referred to as MF theory as far as the exponents are concerned. But it is quantitatively different from the RMA-based mean-field theory in all other respects. It should be noted that a different recursive lattice (e.g., diamond hierarchical) may be used to capture longer-range correlations and generate non-classical critical exponents. It should also be noted that, due to the topology of the Bethe lattice, two polymer chains can never interact at more than one point, producing a situation resembling the single chain contact approximation. However, a different recursive lattice could be chosen (e.g., a Husimi cactus), such that multiple contacts between chains are allowed. In any case, the thermodynamics in our model only depends on the number of contacts between unlike monomers, regardless of how many different chains they belong to. The model has shown the correct behavior of the second virial coefficient. Finally, although the conformation of an isolated polymer chain adsorbed to a surface may be important to a full understanding of our results, our method is only intended to treat infinite systems that cannot be treated by simulations. The monomer density in an infinite system containing only a single polymer chain of finite size is zero at every generation. Therefore, the study of a single finite coil is not possible with our method and requires the use of simulation. An infinitely long single chain with a non-zero monomer density can be easily studied by our approach.

The outline of the paper is as follows. Section I will give an overview of the essential properties of the model. Section II will address the calculations of various densities as a function of distance from the surface. The results of the calculations and conclusions that can be drawn from them are included in Sections III and IV.

## I. THE MODEL

Since an exact calculation of the PF on a cubic lattice is unfeasible, we replace it with a tree structure specially designed to locally emulate a cubic lattice, but allow an exact solution through a recursive calculation technique. The reader is referred to Fig. 1 for a schematic diagram of the structure and to Ref. 6 for a more detailed description. Please note that Fig. 1 is only schematic and cannot show the infinite number of bulk and surface trees in the actual structure. In the original cubic lattice, with an even number of parallel lattice planes between the two surfaces, an imaginary plane parallel to the surfaces and halfway between them bisects a set of central lattice bonds. Looking out from any one of these central bonds toward either surface will produce an identical view. Herein lies the lateral homogeneity and translational invariance of the cubic lattice, which we attempt to emulate with an infinite set of bulk trees in our recursive structure. We imagine each central bond of the cubic lattice transformed into the central bond of a symmetrical Cayley tree, so that the view in either direction from any such bond is identical. A perfect representation of the cubic lattice would require this infinite set of trees to be interconnected within the bulk of the structure; however, this would make the problem insoluble. Instead, we make the strong approximation of replacing the bulk by an infinite set of *independent* bulk trees, connected only through surface sites. Since any bulk tree $\mathfrak{I}$ is finite in extent, and possesses a center, it does not possess in itself the translational invariance and infinite extent that the bulk of the cubic lattice possesses along the directions parallel to the surfaces. However, since the recursive structure as a whole contains an infinite number of replicas of $\Im$, and since a chain can wind its way through multiple replicas, the bulk of the structure is effectively infinite along "directions parallel to the surfaces." A chain in the cubic lattice may ascend from the bulk to one of the two surfaces, then occupy one or more surface sites before descending back into the *same* bulk lattice. A single chain may do this multiple times, resulting in loop structures adsorbed to the surface. An analogous conformation in our recursive structure involves a chain ascending from a given bulk tree into an attached surface tree and then descending into a *different* bulk tree, which may be repeated several times, resulting in the simultaneous occupation of several bulk and surface trees by a single chain. Such a case emulates the "loops" that may adsorb to the surface of the cubic lattice. We should note that, since all the bulk trees are identical in every respect, the solution on any bulk tree is identical to the solution on all trees, so that the homogeneity of the cubic lattice is perfectly maintained.

Each site of the lattice is occupied by one monomer from one of three possible species: the star polymer species A, the linear polymer species B, or the monomeric free volume C (in which case the "monomer" actually represents an unoccupied void site). No site can be occupied by more than one monomer, nor can any polymer chain visit the same site twice. Each species A star polymer consists of one monomer acting as a core, attached to some number of equal-length linear chains acting as the arms. The number of arms in each star is specified by the integer $n$, $0 < n < q,$ where $q$ is the coordination number. The positive integer $N_{\mathrm{A}}$ specifies the number of monomers in each arm of the star. The total number of monomers in the star is therefore given by $1 + N_{\mathrm{A}} n.$ The number of monomers in a species B linear polymer is specified by the positive integer $N_{\mathrm{B}}$. Pairs of monomers on adjacent sites (any pair of sites connected by a lattice bond) interact through a set of excess energies $\varepsilon_{\mu\nu}$ ($\mu,\nu = \mathrm{A,B,C}$), where it is understood that $\varepsilon_{\mu\nu} = \varepsilon_{\nu\mu}$ and $\varepsilon_{\mathrm{AA}} = \varepsilon_{\mathrm{BB}} = \varepsilon_{\mathrm{CC}} = 0$. Therefore, there are only three distinct, nonzero excess energies corresponding to the three possible contacts that can be formed between unlike monomers (AB, AC, BC). Corresponding to the set of excess energies is the set of Boltzmann weights $w_{\mu\nu}$, defined as

$$w_{\mu\nu} \equiv \exp(-\beta\varepsilon_{\mu\nu}),$$

where $\beta \equiv 1/T$, $T$ representing the absolute temperature in the units of the Boltzmann constant, and $w_{\mu\nu} = w_{\nu\mu}$. Note that $w_{\mathrm{AA}} = w_{\mathrm{BB}} = w_{\mathrm{CC}} = 1$. Plate $j = 1,2$ interacts with component $\alpha = \mathrm{A,C}$ with the excess energy $\bar{\varepsilon}_{j,\alpha}$. The B species requires no separate parameter to describe its own interaction with the surface, since the requirement that the densities of all species sum to unity at any site adds a condition to determine the density of B uniquely. This condition is referred to as *the sum rule* in the following. The bulk composition of the film is likewise determined by two activities: the activity $K$ of A species bonds, and the activity $\eta_{\mathrm{C}}$ of C species "monomers" or voids. Hence only two parameters related to A and C are required to specify bulk composition, since the bulk density of B follows from the sum rule. Please note that each interaction energy is only defined to within an arbitrary constant, so that only *differences* in energy are relevant to the calculation. For example, only the *ratios* of Boltzmann weights affect the state of the system. In this sense, surface interaction energies and bulk activities are only relative parameters, arbitrarily adjusted to produce the desired surface and bulk compositions and affecting nothing else, and it would be possible to obtain the same density profile by specifying these parameters for a different pair of components and adjusting their values accordingly. The even integer $D$ specifies the number of lattice generations separating the two plates.

A configuration of the system on a lattice of $N$ sites is uniquely described by the set of parameters $n, N_{\mathrm{A}}, N_{\mathrm{B}}, B_{\mathrm{A}}, N_{\mathrm{C}}, N_{\mathrm{AB}}, N_{\mathrm{AC}}, N_{\mathrm{BC}}, \overline{N}_{j,\mathrm{A}}$, and $\overline{N}_{j,\mathrm{C}}$, where $B_{\mathrm{A}}$ is the total number of A bonds (determined by the activity $K$), $N_{\mathrm{C}}$ is the total number of C monomers (determined by the activity $\eta_{\mathrm{C}}$), $N_{\mu\nu}$ is the total number of contacts between $\mu$- and $\nu$-species monomers, and $\overline{N}_{j,\alpha}$ $(j=1,2,\ \alpha=\mathrm{A},\mathrm{C})$ is the total number of $\alpha$ monomers on the $j$-th surface (determined by the activity $\overline{w}_{j,\alpha}$). The total PF $Z_N$ of the system is

$$Z_N = \sum K^{B_{\mathrm{A}}} \eta_{\mathrm{C}}^{N_{\mathrm{C}}} w_{\mathrm{AB}}^{N_{\mathrm{AB}}} w_{\mathrm{AC}}^{N_{\mathrm{AC}}} w_{\mathrm{BC}}^{N_{\mathrm{BC}}} \overline{w}_{1,\mathrm{A}}^{\overline{N}_{1,\mathrm{A}}} \overline{w}_{2,\mathrm{A}}^{\overline{N}_{2,\mathrm{A}}} \overline{w}_{1,\mathrm{C}}^{\overline{N}_{1,\mathrm{C}}} \overline{w}_{2,\mathrm{C}}^{\overline{N}_{2,\mathrm{C}}},$$

where the summation is taken over all distinct configurations on the lattice. We note that all quantities related to the surfaces are indicated by overbars. The *thermodynamic limit* is obtained by considering the sequence $\{\ln Z_N / N\}$ as $N \to \infty$. In our case, this is done by considering a lattice that is infinitely large in the direction transverse along the two surfaces. Thus, each surface will be eventually an infinitely large surface. All quantities that we report here are in this limit. Henceforth, we will suppress the subscript $N$ in the total PF.

The definitions of various "partial partition functions" (PPF's) and the ratios derived from them, as well as the recursion relations (RR's) relating them, are given in the Appendix, to which the reader is referred for definitions of symbols found in the following. Also given in the Appendix are conventions governing the values of the subscript $i$ that apply to all expressions throughout the rest of this paper.

## II. CALCULATION OF DENSITIES

To calculate the density of a particular feature at an $i$-th site in $\Im_j$, we simply divide the PF of a system constrained to have that feature at the given site by the total PF $Z$. Since the total

PF is just the product of the PPF's of the two pieces generated by cutting an *i*-th bond, multiplied by any weights of interaction between those pieces, it can be calculated by considering all possible states of that bond and its two end sites. The case that it is unoccupied contributes the same set of terms as in the linear-linear blend, $\sum_{\mu,\nu} w_{\mu\nu} Z_{j,i}(1,\mu) Z'_{j,i}(1,\nu)$. If the *i*-th bond is occupied by an A polymer bond, then the sites above and below it may be occupied by an A endpoint of a star arm and the second monomer from the end of that arm, or the second monomer and the third monomer from that end, etc., respectively. We must also take into account that the core of the corresponding star may either be above or below the *i*-th bond. These possibilities generate the set of terms

$\sum_{k=1}^{N_\mathrm{A}} \left[ Z_{j,i}(0,\mathrm{A},\mathrm{e},k) Z'_{j,i}(0,\mathrm{A},\mathrm{c},N_\mathrm{A}-k) + Z_{j,i}(0,\mathrm{A},\mathrm{c},N_\mathrm{A}-k) Z'_{j,i}(0,\mathrm{A},\mathrm{e},k) \right]$. If the *i*-th bond is occupied by a B polymer bond with $N_\mathrm{B} > 1$, the various possible states of the sites above and below it generate the set of terms $\sum_{k=1}^{N_\mathrm{B}-1} Z_{j,i}(0,\mathrm{B},k) Z'_{j,i}(0,\mathrm{B},N_\mathrm{B}-k)$. Hence, $Z$ is just the sum of these three sets of terms, or

$$
\begin{aligned}
Z = {} & \sum_{\mu,\nu} w_{\mu\nu} Z_{j,i}(1,\mu) Z'_{j,i}(1,\nu) + \sum_{k=1}^{N_\mathrm{B}-1} Z_{j,i}(0,\mathrm{B},k) Z'_{j,i}(0,\mathrm{B},N_\mathrm{B}-k) \\
& + \sum_{k=1}^{N_\mathrm{A}} \left[ Z_{j,i}(0,\mathrm{A},\mathrm{e},k) Z'_{j,i}(0,\mathrm{A},\mathrm{c},N_\mathrm{A}-k) + Z_{j,i}(0,\mathrm{A},\mathrm{c},N_\mathrm{A}-k) Z'_{j,i}(0,\mathrm{A},\mathrm{e},k) \right]
\end{aligned}
$$

As an example of the PF of a $\Im$ constrained to have a particular feature at the *i*-th site in $\Im_j$, we will derive that for an A endpoint at the site $0 \le i < I$. The A polymer bond attached to the endpoint may occupy either the *i*-th lattice bond below or one of the *r* (*i*+1)-th lattice bonds above the *i*-th site. In the former case, there is only one possible way to place the polymer bond, and the (*i*+1)-th lattice bonds are all unoccupied and may terminate in any species of monomer.

Therefore, the contribution from this possibility is $K^{1/2} Z'_{j,i}(0,\mathrm{A},\mathrm{c},N_\mathrm{A}-1) U^{r}_{j,i+1}$. In the latter case, there are *r* possible ways to place the polymer bond, and the remaining $r'$ (*i*+1)-th lattice bonds are unoccupied and can terminate in any species of monomer. The contribution from this possibility is $rK^{1/2} U'_{j,i} Z_{j,i+1}(0,\mathrm{A},\mathrm{c},N_\mathrm{A}-1) U^{r'}_{j,i+1}$. Hence, the PF $Z_{j,1,\mathrm{A},i}$ of the system constrained to have an A endpoint at this site is just the sum of the above two contributions, or

$$Z_{j,1,\mathrm{A},i} = K^{1/2} Z'_{j,i}(0,\mathrm{A},\mathrm{c},N_\mathrm{A}-1) U^{r}_{j,i+1} + rK^{1/2} U'_{j,i} Z_{j,i+1}(0,\mathrm{A},\mathrm{c},N_\mathrm{A}-1) U^{r'}_{j,i+1}.$$

At the level $i = I$, the expression is slightly different, namely

$$Z_{j,1,\mathrm{A},I} = \overline{w}_{j,\mathrm{A}} K^{1/2} Z'_{j,I}(0,\mathrm{A},\mathrm{c},N_\mathrm{A}-1) \overline{U}^{r'}_{j,0} + r'\overline{w}_{j,\mathrm{A}} K^{1/2} U'_{j,I} \overline{Z}_{j,0}(0,\mathrm{A},\mathrm{c},N_\mathrm{A}-1) \overline{U}^{r''}_{j,0},$$

The constrained PF's above may be plugged into the expression for endpoint density, which is

$$\phi_{j,1,\mathrm{A},i} = Z_{j,1,\mathrm{A},i} / Z, \quad 0 \le i \le I.$$

Constrained PF's for various other densities of the A species will be given below without derivation. The corresponding quantities for B are given in Ref. 6 and will not be included.

The density $\phi_{j,2,\mathrm{A},i}$ of bifunctional branchings is calculated from

$$\phi_{j,2,\mathrm{A},i} = Z_{j,2,\mathrm{A},i} / Z, \quad 0 \le i \le I, \text{ where}$$

$$Z_{j,2,\mathrm{A},i} = r_2 K U^{r''}_{j,i+1} U'_{j,i} \tilde{Z}^{(\mathrm{c})}_{j,i+1}(0,\mathrm{A}) + rKU^{r'}_{j,i+1} \tilde{Z}'^{(\mathrm{c})}_{j,i+1}(0,\mathrm{A},\mathrm{c}) + rKU^{r'}_{j,i+1} \tilde{Z}'^{(\mathrm{c})}_{j,i+1}(0,\mathrm{A},\mathrm{e}),$$

$$Z_{j,2,\mathrm{A},I} = r'_2 \overline{w}_{j,\mathrm{A}} K \overline{U}^{r'''}_{j,0} U'_{j,I} \overline{\tilde{Z}}^{(\mathrm{c})}_{j,0}(0,\mathrm{A}) + r'\overline{w}_{j,\mathrm{A}} K \overline{U}^{r''}_{j,0} \overline{\tilde{Z}}'^{(\mathrm{c})}_{j,0}(0,\mathrm{A},\mathrm{c}) + r'\overline{w}_{j,\mathrm{A}} K \overline{U}^{r''}_{j,0} \overline{\tilde{Z}}'^{(\mathrm{c})}_{j,0}(0,\mathrm{A},\mathrm{e}).$$

The density of star cores $\phi_{j,c,i}$ at each level is calculated from

$$\phi_{j,c,i} = Z_{j,c,i} / Z, \quad 0 \le i \le I, \text{ where}$$

$$\begin{aligned} Z_{j,c,i} = {} & r_{n-1} K^{n/2} U^{q-n}_{j,i+1} Z^{n-1}_{j,i+1}(0,\mathrm{A},\mathrm{e},N_\mathrm{A}) Z'_{j,i}(0,\mathrm{A},\mathrm{e},N_\mathrm{A}) \\ & + r_n K^{n/2} U^{r-n}_{j,i+1} U'_{j,i} Z^{n}_{j,i+1}(0,\mathrm{A},\mathrm{e},N_\mathrm{A}), \end{aligned}$$

$$Z_{j,c,I} = \begin{Bmatrix} K^{r/2}\overline{Z}_{j,0}^{r'}(0,\mathrm{A},\mathrm{e},N_\mathrm{A})Z'_{j,I}(0,\mathrm{A},\mathrm{e},N_\mathrm{A}), \quad n = r \\ r'_{n-1}K^{n/2}\overline{U}_{j,0}^{r-n}\overline{Z}_{j,0}^{n-1}(0,\mathrm{A},\mathrm{e},N_\mathrm{A})Z'_{j,I}(0,\mathrm{A},\mathrm{e},N_\mathrm{A}) \\ +r'_nK^{n/2}\overline{U}_{j,0}^{r'-n}U'_{j,I}\overline{Z}_{j,0}^{n}(0,\mathrm{A},\mathrm{e},N_\mathrm{A}), \quad n < r \end{Bmatrix}.$$

Once the densities of endpoints, bifunctional branchings, and cores of the star species have all been calculated, the corresponding monomer densities, $\phi_{j,m,\mathrm{A},i}$, are given by

$$\phi_{j,m,\mathrm{A},i} = \phi_{j,1,\mathrm{A},i} + \phi_{j,2,\mathrm{A},i} + \phi_{j,c,i}, \quad 0 \le i \le I.$$

The average numbers of A polymer bonds per site, $\phi_{j,b,\mathrm{A},i}$, are calculated from

$$\phi_{j,b,\mathrm{A},i} = \frac{1}{2}\left(\phi_{j,1,\mathrm{A},i} + 2\phi_{j,2,\mathrm{A},i} + n\phi_{j,c,i}\right), \quad 0 \le i \le I.$$

Since the calculation of contact densities does not depend on polymer architecture, the expressions used to obtain contact densities are the same for star-linear blends as those given for linear-linear blends in Ref. 6. Similarly, the entropy density is calculated by the same method as described in Ref. 6, to which we refer the reader for details.

## III. RESULTS

Of the figures representing the results for star-linear blends, the absolute densities are plotted in some, and the bulk-normalized densities in others. The bulk-normalized density at the *i*-th site in $\Im_j$ is the absolute density at that site divided by its uniform value in the bulk of the tree. We have in all cases treated star-linear blends in which the star component and linear component have different architectures but equal molecular weights. Their common molecular weight is chosen to be as close to 200 monomers as the constraints of star architecture will allow. The bulk compositions of the blends are always fixed at equal densities of star and linear monomers.

In Figs. 2 through 4, we show the bulk-normalized star monomer density as a function of distance from one neutral plate (in units of the end-to-end distance $R_a$ of a star arm) for neutral blends of stars and linear chains with free volume in an essentially semi-infinite geometry. That is, the plate separation is much larger than the correlation lengths of surface segregation at each plate. There are specifically 2000 lattice generations between them. The numbers of lattice generations corresponding to the distances $R_a$ for 3-arm, 4-arm, and 5-arm stars are 66, 50, and 40, respectively. Fig. 2 shows profiles for blends with 3-arm stars, Fig. 3 blends with 4-arm stars, and Fig. 4 blends with 5-arm stars. The value of $\phi_{C,\infty}$ varies from 0% for the solid lines to 30% ($\eta_C = 2.0$) for the dotted lines to 50% ($\eta_C = 3.07$) for the dashed lines. Bulk composition has been set to 50/50 for the 3-arm, 4-arm, and 5-arm blends by choosing $K$ values of 1.006700, 1.016230, and 1.029150, respectively. The insets show the large abrupt drop at the surface in each of the profiles representing compressible systems. As is the case of linear-linear blends, this is an irrelevant feature indicating the enrichment of free volume in the first surface layer, which is caused by the generally known entropic tendency of smaller molecules to segregate to the surface. Since free volume is modeled here as monomeric particles, it constitutes the smallest species in the represented blends. In Figs. 5 through 7, we show the profiles of bulk-normalized free volume density for the same star-linear blends with 3-arm, 4-arm, and 5-arm stars, respectively. The solid curves represent profiles for $\phi_{C,\infty} = 0.30$ and the dotted curves profiles for $\phi_{C,\infty} = 0.50$. The variations in the free volume profiles are extremely small compared to those in the star monomer profiles, allowing us to omit showing the profiles for the linear species monomers as deducible by the reader.

The most prominent feature of all the profiles in Figs. 2 through 4 is the long-range decaying oscillation with a near-surface wavelength approximately twice the end-to-end distance $R_a$ of the star arm. This indicates layering of the star coils at the surface. These oscillations are also seen to increase in amplitude both as $\phi_{C,\infty}$ increases and as $n$ increases. Similar effects are seen in the free volume density profiles in Figs. 5 through 7. These are caused by the growing tendency of the stars to layer themselves next to the surface in either of these cases. It is intuitively clear that increasing $n$ or increasing $\phi_{C,\infty}$ would reduce the tendency of the star coils to interpenetrate. (In our model, Star 1 "penetrates" Star 2 when it occupies one or more of the lattice sites lying within one radius of gyration of the core of Star 2, some of which Star 2 will not occupy.) Another effect of star layering is seen in the shorter-range oscillations in the profiles (shown in the insets), which have a length scale of about one lattice generation. These correspond to similar oscillations on a length scale of one bead diameter in the results of Yethiraj[7].

In all profiles, the linear chains are enriched within a distance of about $\frac{1}{2}R_a$ from the surface, and the stars enriched at a distance $R_a$. This is similar to the results that Yethiraj[7] obtained using integral equation theory. However, his calculations show an enrichment of linear chains only within a few monomer diameters of the surface, and he explains this result as being due to the greater success of packing linear chains against the surface to allow more free volume in the bulk. Our results, by contrast, have a much longer-ranged enrichment of linear chains that exists even when there is no free volume in the mixture. We explain this as a local repulsion of stars from the surface caused by the greater functionality of the core monomers. Since the reduction in coordination number from 6 to 5 at the surface produces a greater loss of entropy to higher functionalities, we would expect higher functionalities to avoid the surface for entropic reasons[8]. However, this local depletion of stars does not indicate the overall enrichment of linear chains. To quantify surface enrichment, we define an integrated surface excess $\sigma_{j,\gamma}$ of the $\gamma$ component in $\Im_j$, graphically represented as the area between the density profile of this component and a horizontal line representing its uniform density in the bulk. For our discrete theory, this can be calculated as the sum of the differences between the local density of that component and its uniform density in the bulk for all lattice generations, i.e.,

$$\sigma_{j,\gamma} = \sum_i (\phi_{j,m,\gamma,i} - \phi_{m,\gamma,\infty}),$$

where $\phi_{j,m,\gamma,i}$ is the monomer density of the $\gamma$ component at the *i*-th lattice generation in $\Im_j$, $\phi_{m,\gamma,\infty}$ is the uniform monomer density of that component in the bulk, and the summation is over all possible values of the generation index *i* in $\Im_j$. This quantity will be positive when the $\gamma$ component is enriched at the surface, and negative when it is depleted. The surface excesses of the star, the linear chain, and the monomeric free volume are designated as $\sigma_\text{A}$, $\sigma_\text{B}$, and $\sigma_\text{C}$, respectively. Their values for each of the blends in Figs. 2 through 4 are listed in Tables I, II, and III, respectively.

The first observation we make concerning these values is that $\sigma_\text{A}$ increases with $\phi_{\text{C},\infty}$ for 3-, 4-, and 5-arm stars. It is also clear that this dependence of $\sigma_\text{A}$ on $\phi_{\text{C},\infty}$ becomes stronger as *n* increases. Since each of the blends is athermal, this must be a purely entropic effect. The loss of entropy suffered by a star as a result of having its coil dimensions constrained by the presence of a hard surface apparently decreases (to a greater degree as *n* is greater) as $\phi_{\text{C},\infty}$ increases. Since a linear chain can be approximated as a 2-arm star, the above supposition implies that its entropy loss at the surface will be reduced by free volume to a lesser degree than any of the stars with which it is mixed. Therefore, the effect of free volume will be different for the two components in a star-linear blend and change their relative entropic attractions to the surface, affecting enrichment to a degree that increases with *n*. Specifically, the tendency of stars to enrich the surface will increase with $\phi_{C,\infty}$, to a greater degree as *n* is greater. As an intuitive explanation for this trend, we suggest that the free volume contained inside a star coil reduces its "hardness," as described in detail in Ref. 6, to a degree that increases with *n*. A linear chain, which can be treated as a 2-arm star, will be softened less than any star polymer with which it is mixed. This explains why the tendency of stars to enrich the surface of a star-linear blend increases with $\phi_{C,\infty}$.

We now move on to studying the dependence of enrichment on *n*. By comparing Tables I-III for $\phi_{C,\infty} = 0$, we see that $\sigma_A$ decreases as *n* increases from 3 to 5 for an incompressible blend. It begins with a positive value for 3-arm stars, and then switches to negative values for greater numbers of arms. This indicates that the star in a blend without free volume is enriched when it has 3 arms, but depleted when it has 4 or 5 arms. Furthermore, the tendency of the star to enrich the surface decreases as *n* grows. This contrasts with the findings of other theoretical works on this topic[7,8], which have found the enrichment of the star to generally increase with *n*. We attribute this apparent discrepancy to the absence of free volume in our own calculation. Our assumption of a fully-packed lattice does not generally hold for calculations of this type performed by other methods. When $\phi_{C,\infty} \neq 0$, the results show a different trend, as will be discussed in the following.

If we compare Tables I-III for $\phi_{\mathrm{C},\infty} = 0.30,$ we see a result that differs in two interesting ways from that of the incompressible case. The more obvious difference is that $\sigma_{\mathrm{A}} > 0$ for both $n = 3$ and $n = 4,$ rather than only for $n = 3$ as for incompressible blends. Hence, the addition of free volume causes the star to be preferred at the surface for a higher number of arms. A more subtle difference lies in the quantitative dependence of $\sigma_{\mathrm{A}}$ on *n*. We see that for $\phi_{\mathrm{C},\infty} = 0,$ $\sigma_{\mathrm{A}}$ monotonically decreases as *n* grows, while for $\phi_{\mathrm{C},\infty} = 0.30,$ it increases from $n = 3$ to $n = 4$ and then decreases from $n = 4$ to $n = 5.$ It is further found by comparing the tables for $\phi_{\mathrm{C},\infty} = 0.50$ that if $\phi_{\mathrm{C},\infty}$ becomes large enough, it eventually reaches a magnitude at which $\sigma_{\mathrm{A}}$ monotonically increases with *n* from 3 to 5. This suggests a general description for the varying trends that we see for different values of $\phi_{\mathrm{C},\infty}$ : for a given value of $\phi_{\mathrm{C},\infty},$ $\sigma_{\mathrm{A}}$ increases with *n* up to a critical arm number $n_{\mathrm{crit}}$, after which it begins to decrease. For $\phi_{\mathrm{C},\infty} = 0,$ $n_{\mathrm{crit}} = 3;$ for $\phi_{\mathrm{C},\infty} = 0.30,$ $n_{\mathrm{crit}} = 4;$ for $\phi_{\mathrm{C},\infty} = 0.50,$ $n_{\mathrm{crit}}$ is an unknown integer satisfying $n_{\mathrm{crit}} \geq 5.$ Our intuitive explanation for this result is as follows.

Two properties of a polymer coil that affect the amount of entropy it loses when it suffers the external constraints of a nearby surface are its hardness and its size. The property of hardness and its role in determining entropy loss is explained in Ref. 6. As stated previously, softer coils will be entropically preferred at the surface. To explain the role of size, we point out that a larger polymer coil suffers greater constraints and loses more entropy than a smaller coil when confined by a rigid wall. This results in an entropic preference for smaller coils at the surface.

As we increase *n* in a star polymer while holding its mass constant, $N_{\mathrm{A}}$ must decrease, along with the size of the coil. This decrease in size taken by itself *increases* the entropic preference for that coil at the surface. However, the increase in the number of arms joined at the core produces more internal constraints to the conformation of the polymer due to excluded volume effects. This increases the hardness of the coil and *reduces* its entropic preference at the surface. There are accordingly two competing effects of increasing *n* on the entropic factors related to surface enrichment. To discuss the interplay of these two effects, we will approximate a linear chain as a star with two arms. We have already seen in Tables I-III (to which we refer the reader in the following) that $\sigma_{\mathrm{A}} > 0$ when $\phi_{\mathrm{C},\infty} = 0$ and $n = 3,$ indicating that stars are enriched at the surface. This implies that as *n* increases from 2 to 3, the entropic preference of a star at the surface also increases; so the size effect predominates in this case. However, as we further increase *n* from 3 to 4, $\sigma_{\mathrm{A}}$ becomes negative, indicating that the hardness effect is now beginning to predominate; the entropic preference of the 4-arm star is lower than that of the 2-arm star. Increasing *n* from 4 to 5 makes $\sigma_{\mathrm{A}}$ even more negative, as the entropic preference for stars at the surface further decreases. Hence, in the absence of free volume, the increasing hardness of a star overcomes its decreasing size when $n > 3.$ That is, $n_{\mathrm{crit}} = 3$ for this case. However, when $\phi_{\mathrm{C},\infty} = 0.30,$ $\sigma_{\mathrm{A}}$ increases as *n* grows from 2 to 4. The reversal of this trend does not occur until $n > 4,$ so that $n_{\mathrm{crit}} = 4$ for this case. The shift in $n_{\mathrm{crit}}$ that results from increased $\phi_{\mathrm{C},\infty}$ can be explained by the softening of the coil by free volume, which has been discussed above. This softening allows the size effect to overcome the hardness effect for higher *n*.

The effect of *n* on the graphs of surface entropy density $S_{\mathrm{S}}$ against $\beta\bar{\varepsilon}_{\mathrm{A}}$ is seen in Fig. 8. The four systems represented are star-linear blends with $w_{\mathrm{AB}} = 1$ and various numbers of arms in the A-species star. The empty triangles, filled circles, empty circles, and filled triangles represent stars with $n = 2,3,4,5,$ respectively. The 2-armed "star" is actually a linear chain. All the systems have $1 + nN_{\mathrm{A}} = N_{\mathrm{B}} = 201,$ except for $n = 3,$ which has $1 + nN_{\mathrm{A}} = N_{\mathrm{B}} = 199.$ As for the linear-linear blends in Fig. 10 of Ref. 6, the graphs in Fig. 8 all have the same asymptotic limit on one side and different asymptotic limits on the other. Since it is the size of the B chain that is identical for these star-linear systems, a common limit exists as $\beta\bar{\varepsilon}_{\mathrm{A}} \to +\infty,$ for which the systems are pure B. The limits as $\beta\bar{\varepsilon}_{\mathrm{A}} \to -\infty$ correspond to pure A and are lower as *n* is higher. This is expected, since a higher functionality at the core produces greater architectural constraints on the possible conformations of a star. The entropy peak shifts further to the right as *n* (and with it the entropy gain associated with enrichment of linear chains) increases. The case of $n = 5$ has no entropy peak, since the entropy of the pure stars is so low that the loss caused by demixing never becomes larger than the gain of linear chain enrichment.

## **IV.** CONCLUSIONS

We have seen that the overall architectural effect of increasing the number of arms in a star-linear blend next to a neutral surface results from the interplay of two competing factors: the decrease in size of the star arms/coil as the number of arms increases while the total number of monomers remains constant; and the increase in the "hardness" of the star coil resulting from packing the same number of segments into its smaller volume. As explained previously, this "hardness" is an intuitive picture expressing the lower configurational entropy of a more compact structure. The decrease in the size of the coil produces an entropic attraction to the surface, while the increase in hardness produces an entropic repulsion. The latter effect predominates for the case of an incompressible system. The integrated surface excess of star segments decreases with increasing number of arms through the entire range of branching functionalities studied here. However, the addition of free volume “softens” the star coils by increasing their configurational entropy, reducing the entropic repulsion and allowing the effect of smaller size to predominate under certain conditions. For a sufficient bulk concentration of free volume, the star excess is seen to increase with the number of arms up to a certain critical arm number, after which it begins to decrease. This critical number increases with the bulk concentration of free volume, indicating a decreasing influence of the hardness aspect of increasing arm number on the overall segregation.

APPENDIX

Cutting an $i$-th bond in $\Im_j$ separates $\Im$ into two parts: the part $\Im'_{j,i}$ that contains the bond $i = 0,$ and the remaining part $\Im_{j,i}$. We define one PF of each part ("partial partition function", PPF) under each possible condition of the cut lattice bond and its end site within that part. In this particular application of the Gujrati-Chhajer approach, we define the following PPF's:

1) $Z_{j,i}(1,\gamma)$ is the PPF of a $\Im_{j,i}$ cut from an unoccupied bond that terminates in a $\gamma$-species monomer.

2) $Z_{j,i}(0,\mathrm{B},k)$ is the PPF of a $\Im_{j,i}$ that is cut from a bond occupied by a B polymer bond, and that contains $k$ monomers from the corresponding chain.

3) $Z_{j,i}(0,\mathrm{A},\mathrm{e},k),\ \ 1 \le k \le N_{\mathrm{A}}, 0 \le i \le I,$ is the PPF of a $\Im_{j,i}$ that is cut from an A species polymer bond in a star arm, and that contains $k$ monomers from this arm, including its endpoint.

4) $Z_{j,i}(0,\mathrm{A},\mathrm{c},k),\ \ 0 \le k \le N_{\mathrm{A}} - 1, 0 \le i \le I,$ is the PPF of a $\Im_{j,i}$ that is cut from an A species polymer bond in a star arm, and that contains $k$ monomers from this arm in addition to the core of the corresponding star.

The PPF's of $\Im'_{j,i}$ are defined and represented in a similar way, except that they are distinguished by the identity of the monomer "below" rather than "above" the cut bond, and by a prime in the notation. PPF's are also defined similarly for the sections of $\overline{\Im}_j,$ which can likewise be cut at an $i$-th bond to yield $\overline{\Im}_{j,i}$ and $\overline{\Im}'_{j,i}$. The representations of these PPF's are the same as those of the others, except for the inclusion of an overhead bar.

Example derivations of the RR's between various PPF's are given in previous papers[5,6], to which we refer the reader for details. Alternatively, anyone with further questions is invited to contact the authors for clarification. The RR's are presented below without derivation.

We find it convenient to introduce the following combinations $U_{j,i}$, $V_{j,i}$, and $W_{j,i}$

$$U_{j,i} \equiv Z_{j,i}(1,\mathrm{A}) + w_{\mathrm{AB}} Z_{j,i}(1,\mathrm{B}) + w_{\mathrm{AC}} Z_{j,i}(1,\mathrm{C}),$$

$$V_{j,i} \equiv Z_{j,i}(1,\mathrm{B}) + w_{\mathrm{AB}} Z_{j,i}(1,\mathrm{A}) + w_{\mathrm{BC}} Z_{j,i}(1,\mathrm{C}),$$

$$W_{j,i} \equiv Z_{j,i}(1,\mathrm{C}) + w_{\mathrm{AC}} Z_{j,i}(1,\mathrm{A}) + w_{\mathrm{BC}} Z_{j,i}(1,\mathrm{B}).$$

We similarly define $U'_{j,i}$, $V'_{j,i}$, and $W'_{j,i}$ by replacing unprimed PPFs by primed PPFs, and $\overline{U}_{j,i}$, $\overline{V}_{j,i}$, and $\overline{W}_{j,i}$ by replacing unprimed PPFs by barred PPFs. We also find it convenient to introduce the following quantities to simplify the notation:

$$r \equiv q-1,\ \ r' \equiv q-2,\ \ r'' \equiv q-3,\ \ r''' \equiv q-4,\ \ r'''' \equiv q-5,$$

$$r_k \equiv \binom{r}{k},\ \ r'_k \equiv \binom{r'}{k},\ \ r''_k \equiv \binom{r''}{k},\ \ r'''_k \equiv \binom{r'''}{k},\ \ r''''_k \equiv \binom{r''''}{k}.$$

$$\tilde{Z}^{(\mathrm{c})}_{j,i}(0,\mathrm{A}) \equiv \sum_{k=1}^{N_{\mathrm{A}}-1} Z_{j,i}(0,\mathrm{A},\mathrm{e},k) Z_{j,i}(0,\mathrm{A},\mathrm{c},N_{\mathrm{A}}-k-1),$$

$$\tilde{Z}'^{(\mathrm{c})}_{j,i}(0,\mathrm{A},\mathrm{c}) \equiv \sum_{k=1}^{N_{\mathrm{A}}-1} Z_{j,i}(0,\mathrm{A},\mathrm{e},k) Z'_{j,i-1}(0,\mathrm{A},\mathrm{c},N_{\mathrm{A}}-k-1),$$

$$\tilde{Z}'^{(\mathrm{c})}_{j,i}(0,\mathrm{A},\mathrm{e}) \equiv \sum_{k=1}^{N_{\mathrm{A}}-1} Z'_{j,i-1}(0,\mathrm{A},\mathrm{e},k) Z_{j,i}(0,\mathrm{A},\mathrm{c},N_{\mathrm{A}}-k-1),$$

$$\overline{\tilde{Z}}^{(\mathrm{c})}_{j,i}(0,\mathrm{A}) \equiv \sum_{k=1}^{N_{\mathrm{A}}-1} \overline{Z}_{j,i}(0,\mathrm{A},\mathrm{e},k) \overline{Z}_{j,i}(0,\mathrm{A},\mathrm{c},N_{\mathrm{A}}-k-1),$$

$$\overline{\tilde{Z}}'^{(\mathrm{c})}_{j,i}(0,\mathrm{A},\mathrm{c}) \equiv \sum_{k=1}^{N_{\mathrm{A}}-1} \overline{Z}_{j,i}(0,\mathrm{A},\mathrm{e},k) Z'_{j,I}(0,\mathrm{A},\mathrm{c},N_{\mathrm{A}}-k-1),$$

$$\bar{\bar{Z}}'^{(\mathrm{c})}_{j,i}(0,\mathrm{A},\mathrm{e}) \equiv \sum_{k=1}^{N_\mathrm{A}-1} Z'_{j,I}(0,\mathrm{A},\mathrm{e},k)\bar{Z}_{j,i}(0,\mathrm{A},\mathrm{c},N_\mathrm{A}-k-1).$$

(Please note that, when $N_\mathrm{A} = 1$, none of the above summations exists, and each will be taken to equal zero. Also, the superscript c should not be confused with species C in the model.) All the RR's used to calculate PPF's that are labeled as referring to species B or C are the same as those given in Ref. 6 for linear-linear blends. They will not be presented below. It is helpful to compare the RR's used to calculate PPF's that are labeled as referring to the star species A with those used to calculate PPF's referring to the linear species B. The presence of the core in a star makes possible two distinguishable orientations of a star arm about one of its bonds: the core can either be "above" or "below" the bond. This adds a new parameter to the star PPF's that produces a larger number of RR's for A than for B. However, a linear chain can be considered as a star containing only two arms and an odd number of monomers. Therefore, the RR's for B are very similar in form to those obtained for A under the same conditions (including $n = 2$).

We further establish the following conventions for the subscript *i* that apply to all the expressions in this paper. When attached to a barred quantity, it may represent any non-negative integer. When used with an unbarred quantity that is a density or is being defined explicitly as a ratio, it represents an integer within the limits $0 \le i \le I$ unless otherwise indicated. With an unbarred, primed quantity, $0 < i \le I$. In all other instances, $0 \le i < I$ unless otherwise indicated.

The RR's for the PPF's of any $\Im_{j,i}$ are given by

$$Z_{j,i}(0,\mathrm{A},\mathrm{e},k) = \begin{Bmatrix} K^{1/2}U^{r}_{j,i+1}, \quad k=1 \\ rKZ_{j,i+1}(0,\mathrm{A},\mathrm{e},k-1)U^{r'}_{j,i+1}, \quad 1 < k \le N_\mathrm{A} \end{Bmatrix},$$

$$Z_{j,i}(0,\mathrm{A},\mathrm{c},k)=\left\{\begin{matrix} r_{n-1}K^{n/2}Z^{n-1}_{j,i+1}(0,\mathrm{A},\mathrm{e},N_\mathrm{A})U^{q-n}_{j,i+1}, & k=0 \\ rKZ_{j,i+1}(0,\mathrm{A},\mathrm{c},k-1)U^{r'}_{j,i+1}, & 0<k<N_\mathrm{A} \end{matrix}\right\},$$

$$Z_{j,i}(1,\mathrm{A})=\left\{\begin{matrix} \Xi_{j,i+1,1}, & N_\mathrm{A}=1 \\ \Xi_{j,i+1,N_\mathrm{A}}+rr'K\tilde{Z}^{(\mathrm{c})}_{j,i+1}(0,\mathrm{A})U^{r''}_{j,i+1}, & N_\mathrm{A}>1 \end{matrix}\right\},$$

where $\Xi_{j,i,\kappa}\equiv rK^{1/2}Z_{j,i}(0,\mathrm{A},\mathrm{c},\kappa-1)U^{r'}_{j,i}+r_nK^{n/2}Z^{n}_{j,i}(0,\mathrm{A},\mathrm{e},\kappa)U^{r-n}_{j,i}$.

The RR's for the PPF's of any $\Im'_{j,i}$ are given by

$$Z'_{j,i}(0,\mathrm{A},\mathrm{e},k)=\left\{\begin{matrix} K^{1/2}U^{r'}_{j,i}U'_{j,i-1}, \quad k=1 \\ r'KZ_{j,i}(0,\mathrm{A},\mathrm{e},k-1)U^{r''}_{j,i}U'_{j,i-1} \\ \quad +KZ'_{j,i-1}(0,\mathrm{A},\mathrm{e},k-1)U^{r'}_{j,i}, \quad 1<k\le N_\mathrm{A} \end{matrix}\right\},$$

$$Z'_{j,i}(0,\mathrm{A},\mathrm{c},k)=\left\{\begin{matrix} r'_{n-2}K^{n/2}Z^{n-2}_{j,i}(0,\mathrm{A},\mathrm{e},N_\mathrm{A})Z'_{j,i-1}(0,\mathrm{A},\mathrm{e},N_\mathrm{A})U^{q-n}_{j,i} \\ \quad +r'_{n-1}K^{n/2}Z^{n-1}_{j,i}(0,\mathrm{A},\mathrm{e},N_\mathrm{A})U^{r-n}_{j,i}U'_{j,i-1}, \quad k=0 \\ r'KZ_{j,i}(0,\mathrm{A},\mathrm{c},k-1)U^{r''}_{j,i}U'_{j,i-1} \\ \quad +KZ'_{j,i-1}(0,\mathrm{A},\mathrm{c},k-1)U^{r'}_{j,i}, \quad 0<k<N_\mathrm{A} \end{matrix}\right\},$$

$$Z'_{j,i}(1,\mathrm{A})=\left\{\begin{matrix} \Xi''_{j,i,0}+\Xi'_{j,i,1}, \quad N_\mathrm{A}=1, n=r \\ \Xi''_{j,i,0}+\Xi'''_{j,i,1}+\Xi''''_{j,i,1}, \quad N_\mathrm{A}=1, n<r \\ \Xi''_{j,i,N_\mathrm{A}-1}+\tilde{\Xi}''_{j,i}+\Xi'_{j,i,N_\mathrm{A}}, \quad N_\mathrm{A}>1, n=r \\ \Xi''_{j,i,N_\mathrm{A}-1}+\tilde{\Xi}''_{j,i}+\Xi'''_{j,i,N_\mathrm{A}}+\Xi''''_{j,i,N_\mathrm{A}}, \quad N_\mathrm{A}>1, n<r \end{matrix}\right\},$$

where $\Xi''_{j,i,\kappa}\equiv r'K^{1/2}Z_{j,i}(0,\mathrm{A},\mathrm{c},\kappa)U^{r''}_{j,i}U'_{j,i-1}+K^{1/2}Z'_{j,i-1}(0,\mathrm{A},\mathrm{c},\kappa)U^{r'}_{j,i}$,

$\tilde{\Xi}''_{j,i}\equiv r'r''K\tilde{Z}^{(\mathrm{c})}_{j,i}(0,\mathrm{A})U^{r'''}_{j,i}U'_{j,i-1}+r'K\tilde{Z}'^{(\mathrm{c})}_{j,i}(0,\mathrm{A},\mathrm{c})U^{r''}_{j,i}+r'K\tilde{Z}'^{(\mathrm{c})}_{j,i}(0,\mathrm{A},\mathrm{e})U^{r''}_{j,i}$,

$\Xi'_{j,i,\kappa}\equiv K^{r/2}Z^{r'}_{j,i}(0,\mathrm{A},\mathrm{e},\kappa)Z'_{j,i-1}(0,\mathrm{A},\mathrm{e},\kappa)$,

$\Xi'''_{j,i,\kappa}\equiv r'_nK^{n/2}Z^{n}_{j,i}(0,\mathrm{A},\mathrm{e},\kappa)U^{r'-n}_{j,i}U'_{j,i-1}$,

$\Xi''''_{j,i,\kappa}\equiv r'_{n-1}K^{n/2}Z^{n-1}_{j,i}(0,\mathrm{A},\mathrm{e},\kappa)Z'_{j,i-1}(0,\mathrm{A},\mathrm{e},\kappa)U^{r-n}_{j,i}$.

The RR's for the PPF's of any $\overline{\Im}_{j,i}$ are given by

$$\overline{Z}_{j,i}(0,\mathrm{A},\mathrm{e},k)=\left\{\begin{array}{l}\overline{w}_{j,\mathrm{A}}K^{1/2}\overline{U}_{j,i+1}^{r''}U'_{j,I}, \quad k=1\\ r''\overline{w}_{j,\mathrm{A}}K\overline{Z}_{j,i+1}(0,\mathrm{A},\mathrm{e},k-1)\overline{U}_{j,i+1}^{r'''}U'_{j,I}\\ \quad +\overline{w}_{j,\mathrm{A}}KZ'_{j,I}(0,\mathrm{A},\mathrm{e},k-1)\overline{U}_{j,i+1}^{r''}, \quad 1<k\le N_{\mathrm{A}}\end{array}\right\},$$

$$\overline{Z}_{j,i}(0,\mathrm{A},\mathrm{c},k)=\left\{\begin{array}{l}\overline{\Xi}_{j,i+1}, \quad k=0, n=r\\ \overline{\Xi}_{j,i+1}+r''_{n-1}\overline{w}_{j,\mathrm{A}}K^{n/2}\overline{Z}_{j,i+1}^{n-1}(0,\mathrm{A},\mathrm{e},N_{\mathrm{A}})\overline{U}_{j,i+1}^{r'-n}U'_{j,I}, \quad k=0, n<r\\ r''\overline{w}_{j,\mathrm{A}}K\overline{Z}_{j,i+1}(0,\mathrm{A},\mathrm{c},k-1)\overline{U}_{j,i+1}^{r'''}U'_{j,I}\\ \quad +\overline{w}_{j,\mathrm{A}}KZ'_{j,I}(0,\mathrm{A},\mathrm{c},k-1)\overline{U}_{j,i+1}^{r''}, \quad 0<k<N_{A}\end{array}\right\},$$

where $\overline{\Xi}_{j,i}\equiv r''_{n-2}\overline{w}_{j,\mathrm{A}}K^{n/2}\overline{Z}_{j,i}^{n-2}(0,\mathrm{A},\mathrm{e},N_{\mathrm{A}})\overline{U}_{j,i}^{r-n}Z'_{j,I}(0,\mathrm{A},\mathrm{e},N_{\mathrm{A}})$,

$$\overline{Z}_{j,i}(1,\mathrm{A})=\left\{\begin{array}{l}\overline{\Xi}_{j,i+1,0}, \quad N_{\mathrm{A}}=1, n=r\\ \overline{\Xi}_{j,i+1,0}+\overline{\Xi}'_{j,i+1,1}, \quad N_{\mathrm{A}}=1, n=r'\\ \overline{\Xi}_{j,i+1,0}+\overline{\Xi}''_{j,i+1,1}+\overline{\Xi}'''_{j,i+1,1}, \quad N_{\mathrm{A}}=1, n<r'\\ \overline{\Xi}_{j,i+1,N_{\mathrm{A}}-1}+\tilde{\overline{\Xi}}_{j,i+1}, \quad N_{\mathrm{A}}>1, n=r\\ \overline{\Xi}_{j,i+1,N_{\mathrm{A}}-1}+\overline{\Xi}'_{j,i+1,N_{\mathrm{A}}}+\tilde{\overline{\Xi}}_{j,i+1}, \quad N_{\mathrm{A}}>1, n=r'\\ \overline{\Xi}_{j,i+1,N_{\mathrm{A}}-1}+\overline{\Xi}''_{j,i+1,N_{\mathrm{A}}}+\overline{\Xi}'''_{j,i+1,N_{\mathrm{A}}}+\tilde{\overline{\Xi}}_{j,i+1}, \quad N_{\mathrm{A}}>1, n<r'\end{array}\right\},$$

where $\overline{\Xi}_{j,i,\kappa}\equiv\overline{w}_{j,\mathrm{A}}K^{1/2}\overline{U}_{j,i}^{r'''}\left[r''\overline{Z}_{j,i}(0,\mathrm{A},\mathrm{c},\kappa)U'_{j,I}+Z'_{j,I}(0,\mathrm{A},\mathrm{c},\kappa)\overline{U}_{j,i}\right]$,

$$\overline{\Xi}'_{j,i,\kappa}\equiv\overline{w}_{j,\mathrm{A}}K^{r'/2}\overline{Z}_{j,i}^{r''}(0,\mathrm{A},\mathrm{e},\kappa)Z'_{j,I}(0,\mathrm{A},\mathrm{e},\kappa),$$

$$\overline{\Xi}''_{j,i,\kappa}\equiv r''_{n}\overline{w}_{j,\mathrm{A}}K^{n/2}\overline{Z}_{j,i}^{n}(0,\mathrm{A},\mathrm{e},\kappa)\overline{U}_{j,i}^{r''-n}U'_{j,I},$$

$$\overline{\Xi}'''_{j,i,\kappa}\equiv r''_{n-1}\overline{w}_{j,\mathrm{A}}K^{n/2}\overline{Z}_{j,i}^{n-1}(0,\mathrm{A},\mathrm{e},\kappa)Z'_{j,I}(0,\mathrm{A},\mathrm{e},\kappa)\overline{U}_{j,i}^{r'-n},$$

$$\tilde{\overline{\Xi}}_{j,i}\equiv r''\overline{w}_{j,\mathrm{A}}K\tilde{\overline{Z}}'^{(\mathrm{c})}_{j,i}(0,\mathrm{A},\mathrm{e})\overline{U}_{j,i}^{r'''}+r''\overline{w}_{j,\mathrm{A}}K\tilde{\overline{Z}}'^{(\mathrm{c})}_{j,i}(0,\mathrm{A},\mathrm{c})\overline{U}_{j,i}^{r'''}$$
$$+r''r'''\overline{w}_{j,\mathrm{A}}K\tilde{\overline{Z}}^{(\mathrm{c})}_{j,i}(0,\mathrm{A})\overline{U}_{j,i}^{r''''}U'_{j,I}.$$

The RR's for the PPF's of any $\Im_{j,I}$ are given by

$$Z_{j,I}(0,\mathrm{A},\mathrm{e},k)=\begin{Bmatrix} \overline{w}_{j,A}K^{1/2}\overline{U}_{j,0}^{r'}, & k=1 \\ r'\overline{w}_{j,\mathrm{A}}K\overline{Z}_{j,0}(0,\mathrm{A},\mathrm{e},k-1)\overline{U}_{j,0}^{r''}, & 1<k\le N_A \end{Bmatrix},$$

$$Z_{j,I}(0,\mathrm{A},\mathrm{c},k)=\begin{Bmatrix} r'_{n-1}\overline{w}_{j,\mathrm{A}}K^{n/2}\overline{Z}_{j,0}^{n-1}(0,\mathrm{A},\mathrm{e},N_\mathrm{A})\overline{U}_{j,0}^{r-n}, & k=0 \\ r'\overline{w}_{j,\mathrm{A}}K\overline{Z}_{j,0}(0,\mathrm{A},\mathrm{c},k-1)\overline{U}_{j,0}^{r''}, & 0<k<N_\mathrm{A} \end{Bmatrix},$$

$$Z_{j,I}(1,\mathrm{A})=\begin{Bmatrix} \Psi_{j,0}, & N_\mathrm{A}=1, n=r \\ \Psi_{j,0}+\overline{\Psi}_{j,1}, & N_\mathrm{A}=1, n<r \\ \Psi_{j,N_\mathrm{A}-1}+\tilde{\Psi}_j, & N_\mathrm{A}>1, n=r \\ \Psi_{j,N_\mathrm{A}-1}+\overline{\Psi}_{j,N_\mathrm{A}}+\tilde{\Psi}_j, & N_\mathrm{A}>1, n<r \end{Bmatrix},$$

where $\Psi_{j,\kappa}\equiv r'\overline{w}_{j,\mathrm{A}}K^{1/2}\overline{Z}_{j,0}(0,\mathrm{A},\mathrm{c},\kappa)\overline{U}_{j,0}^{r''}$,

$$\tilde{\Psi}_j\equiv r'r''\overline{w}_{j,\mathrm{A}}K\tilde{\overline{Z}}_{j,0}^{(\mathrm{c})}(0,\mathrm{A})\overline{U}_{j,0}^{r'''},$$

$$\overline{\Psi}_{j,\kappa}\equiv r'_n\overline{w}_{j,\mathrm{A}}K^{n/2}\overline{Z}_{j,0}^{n}(0,\mathrm{A},\mathrm{e},\kappa)\overline{U}_{j,0}^{r'-n}.$$

The RR's for the PPF's of any $\Im'_{j,0}$ are given by

$$\overline{Z}'_{j,0}(0,\mathrm{A},\mathrm{e},k)=\begin{Bmatrix} \overline{w}_{j,\mathrm{A}}K^{1/2}\overline{U}_{j,0}^{r''}U'_{j,I}, & k=1 \\ r''\overline{w}_{j,\mathrm{A}}K\overline{Z}_{j,0}(0,\mathrm{A},\mathrm{e},k-1)\overline{U}_{j,0}^{r'''}U'_{j,I} & \\ \quad +\overline{w}_{j,\mathrm{A}}KZ'_{j,I}(0,\mathrm{A},\mathrm{e},k-1)\overline{U}_{j,0}^{r''}, & 1<k\le N_\mathrm{A} \end{Bmatrix},$$

$$\overline{Z}'_{j,0}(0,\mathrm{A},\mathrm{c},k)=\begin{Bmatrix} \overline{w}_{j,\mathrm{A}}K^{r/2}\overline{Z}_{j,0}^{r''}(0,\mathrm{A},\mathrm{e},N_\mathrm{A})Z'_{j,I}(0,\mathrm{A},\mathrm{e},N_\mathrm{A}), & k=0, n=r \\ r''_{n-1}\overline{w}_{j,\mathrm{A}}K^{n/2}\overline{Z}_{j,0}^{n-1}(0,\mathrm{A},\mathrm{e},N_\mathrm{A})\overline{U}_{j,0}^{r'-n}U'_{j,I} & \\ \quad +r''_{n-2}\overline{w}_{j,\mathrm{A}}K^{n/2}\overline{Z}_{j,0}^{n-2}(0,\mathrm{A},\mathrm{e},N_\mathrm{A})Z'_{j,I}(0,\mathrm{A},\mathrm{e},N_\mathrm{A})\overline{U}_{j,0}^{r-n}, & k=0, n<r \\ r''\overline{w}_{j,\mathrm{A}}K\overline{Z}_{j,0}(0,\mathrm{A},\mathrm{c},k-1)\overline{U}_{j,0}^{r'''}U'_{j,I} & \\ \quad +\overline{w}_{j,\mathrm{A}}KZ'_{j,I}(0,\mathrm{A},\mathrm{c},k-1)\overline{U}_{j,0}^{r''}, & 0<k<N_\mathrm{A} \end{Bmatrix},$$

$$
\bar{Z}'_{\alpha,0}(1,\mathrm{A}) = \left\{ \begin{array}{l}
\bar{\Psi}'_{j,0}, \quad N_{\mathrm{A}} = 1, n = r \\
\bar{\Psi}'_{j,0} + \bar{\Xi}'_{j,0,1}, \quad N_{\mathrm{A}} = 1, n = r' \\
\bar{\Psi}'_{j,0} + \bar{\Xi}''_{j,0,1} + \bar{\Xi}'''_{j,0,1}, \quad N_{\mathrm{A}} = 1, n < r' \\
\bar{\Psi}'_{j,N_{\mathrm{A}}-1} + \tilde{\tilde{\Xi}}_{j,0}, \quad N_{\mathrm{A}} > 1, n = r \\
\bar{\Psi}'_{j,N_{\mathrm{A}}-1} + \tilde{\tilde{\Xi}}_{j,0} + \bar{\Xi}'_{j,0,N_{\mathrm{A}}}, \quad N_{\mathrm{A}} > 1, n = r' \\
\bar{\Psi}'_{j,N_{\mathrm{A}}-1} + \bar{\Xi}'''_{j,0,N_{\mathrm{A}}} + \bar{\Xi}''_{j,0,N_{\mathrm{A}}} + \tilde{\tilde{\Xi}}_{j,0}, \quad N_{\mathrm{A}} > 1, n < r'
\end{array} \right\},
$$

where $\bar{\Psi}'_{j,\kappa} \equiv \bar{w}_{j,\mathrm{A}} K^{1/2} \left[ r'' \bar{Z}_{j,0}(0,\mathrm{A},\mathrm{c},\kappa) \bar{U}_{j,0}^{r'''} U'_{j,I} + Z'_{j,I}(0,\mathrm{A},\mathrm{c},\kappa) \bar{U}_{j,0}^{r''} \right],$

As in the case of linear-linear blends**,** each of the PPF's grows exponentially with iteration, so it is necessary for the calculation method to define ratios of the PPF's, by dividing each one by some function of them. These PPF ratios are then used in the calculations by deriving RR's between them from the corresponding relations between PPF's. The definitions of the ratios labeled as referring to species A and their corresponding RR's are given below, whenever they are different from those given in Ref. 6 for linear-linear blends. The RR's are easily obtained from those given above for the PPF's, and are shown below without derivation. Once again, the definitions of ratios labeled as referring to species B or C, and any such quantities defined for simplicity of notation, are identical to the corresponding expressions given in Ref. 6. They will not be included in the Appendix. The RR's between the ratios are used in an iterative calculation to determine the value of each of the ratios at each bond of $\Im$, as described in detail in previous papers by this group[1-5]. It is then trivial to use these ratios to calculate the total PF of the lattice, and from this the density at each lattice generation of any species of monomer, endpoint, inter-monomer contact, etc., that might interest us. (The equations used to calculate the various densities in terms of the ratios will be presented below.) It also allows an easy calculation of free energy and entropy densities of $\Im$, which are difficult to obtain in MC simulations.

The ratios are defined as follows:

$$y_{j,\mathrm{A,e},k,i} \equiv Z_{j,i}(0,\mathrm{A},\mathrm{e},k)\big/U_{j,i}\,, \qquad y_{j,\mathrm{A,c},k,i} \equiv Z_{j,i}(0,\mathrm{A},\mathrm{c},k)\big/U_{j,i}\,,$$

$$y'_{j,\mathrm{A,e},k,i} \equiv Z'_{j,i}(0,\mathrm{A},\mathrm{e},k)\big/U'_{j,i}\,, \qquad y'_{j,\mathrm{A,c},k,i} \equiv Z'_{j,i}(0,\mathrm{A},\mathrm{c},k)\big/U'_{j,i}\,,$$

$$\bar{y}_{j,\mathrm{A,e},k,i} \equiv \bar{Z}_{j,i}(0,\mathrm{A},\mathrm{e},k)\big/\bar{U}_{j,i}\,, \qquad \bar{y}_{j,\mathrm{A,c},k,i} \equiv \bar{Z}_{j,i}(0,\mathrm{A},\mathrm{c},k)\big/\bar{U}_{j,i}\,,$$

The notation in the equations relating these ratios is simplified by the following definitions:

$$\tilde{y}^{(\mathrm{c})}_{j,\mathrm{A},i} \equiv \sum_{k=1}^{N_\mathrm{A}-1} y_{j,\mathrm{A,e},k,i}\, y_{j,\mathrm{A,c},N_\mathrm{A}-k-1,i}\,, \qquad \tilde{y}'^{(\mathrm{c})}_{j,\mathrm{A,c},i} \equiv \sum_{k=1}^{N_\mathrm{A}-1} y_{j,\mathrm{A,e},k,i}\, y'_{j,\mathrm{A,c},N_\mathrm{A}-k-1,i-1}\,,$$

$$\tilde{y}'^{(\mathrm{c})}_{j,\mathrm{A,e},i} \equiv \sum_{k=1}^{N_\mathrm{A}-1} y'_{j,\mathrm{A,e},k,i-1}\, y_{j,\mathrm{A,c},N_\mathrm{A}-k-1,i}\,, \qquad \tilde{\bar{y}}^{(\mathrm{c})}_{j,\mathrm{A},i} \equiv \sum_{k=1}^{N_\mathrm{A}-1} \bar{y}_{j,\mathrm{A,e},k,i}\, \bar{y}_{j,\mathrm{A,c},N_\mathrm{A}-k-1,i}\,,$$

$$\tilde{\bar{y}}'^{(\mathrm{c})}_{j,\mathrm{A,c},i} \equiv \sum_{k=1}^{N_\mathrm{A}-1} \bar{y}_{j,\mathrm{A,e},k,i}\, y'_{j,\mathrm{A,c},N_\mathrm{A}-k-1,I}\,, \qquad \tilde{\bar{y}}'^{(\mathrm{c})}_{j,\mathrm{A,e},i} \equiv \sum_{k=1}^{N_\mathrm{A}-1} y'_{j,\mathrm{A,e},k,I}\, \bar{y}_{j,\mathrm{A,c},N_\mathrm{A}-k-1,i}\,.$$

(Please note that, when $N_\mathrm{A} = 1,$ none of the above summations exists, and each will be taken to equal zero. Also, the superscript c should not be confused with species C in the model.)

$$Y_{j,\mathrm{A,e},k,i} \equiv \begin{Bmatrix} K^{1/2}, & k = 1 \\ rKy_{j,\mathrm{A,e},k-1,i+1}, & 1 < k \le N_\mathrm{A} \end{Bmatrix},$$

$$Y_{j,\mathrm{A,c},k,i} \equiv \begin{Bmatrix} r_{n-1}K^{n/2} y^{n-1}_{j,\mathrm{A,e},N_\mathrm{A},i+1}, & k = 0 \\ rKy_{j,\mathrm{A,c},k-1,i+1}, & 0 < k < N_\mathrm{A} \end{Bmatrix},$$

$$X_{j,\mathrm{A},i} \equiv \begin{Bmatrix} \xi_{j,i+1,1}, & N_\mathrm{A} = 1 \\ \xi_{j,i+1,N_\mathrm{A}} + rr'K\tilde{y}^{(\mathrm{c})}_{j,\mathrm{A},i+1}, & N_\mathrm{A} > 1 \end{Bmatrix},$$

where $\xi_{j,i,\kappa} \equiv rK^{1/2} y_{j,\mathrm{A,c},\kappa-1,i} + r_n K^{n/2} y^{n}_{j,\mathrm{A,e},\kappa,i}\,,$

$$Y'_{j,\mathrm{A,e},k,i} \equiv \begin{Bmatrix} K^{1/2}, & k = 1 \\ r'Ky_{j,\mathrm{A,e},k-1,i} + Ky'_{j,\mathrm{A,e},k-1,i-1}, & 1 < k \le N_\mathrm{A} \end{Bmatrix},$$

$$Y'_{j,\mathrm{A,c},k,i} \equiv \begin{Bmatrix} r'_{n-2}K^{n/2}y^{n-2}_{j,\mathrm{A,e},N_\mathrm{A},i}y'_{j,\mathrm{A,e},N_\mathrm{A},i-1} + r'_{n-1}K^{n/2}y^{n-1}_{j,\mathrm{A,e},N_\mathrm{A},i}, & k = 0 \\ r'Ky_{j,\mathrm{A,c},k-1,i} + Ky'_{j,\mathrm{A,c},k-1,i-1}, & 0 < k < N_\mathrm{A} \end{Bmatrix},$$

$$X'_{j,\mathrm{A},i} \equiv \begin{Bmatrix} \xi''_{j,i,0} + \xi'_{j,i,1}, & N_\mathrm{A} = 1, n = r \\ \xi''_{j,i,0} + \xi'''_{j,i,1} + \xi''''_{j,i,1}, & N_\mathrm{A} = 1, n < r \\ \xi''_{j,i,N_\mathrm{A}-1} + \tilde{\xi}''_{j,i} + \xi'_{j,i,N_\mathrm{A}}, & N_\mathrm{A} > 1, n = r \\ \xi''_{j,i,N_\mathrm{A}-1} + \tilde{\xi}''_{j,i} + \xi'''_{j,i,N_\mathrm{A}} + \xi''''_{j,i,N_\mathrm{A}}, & N_\mathrm{A} > 1, n < r \end{Bmatrix},$$

where $\xi''_{j,i,\kappa} \equiv r'K^{1/2}y_{j,\mathrm{A,c},\kappa,i} + K^{1/2}y'_{j,\mathrm{A,c},\kappa,i-1}$,

$\xi'_{j,i,\kappa} = K^{r/2}y^{r'}_{j,\mathrm{A,e},\kappa,i}y'_{j,\mathrm{A,e},\kappa,i-1}$,

$\xi'''_{j,i,\kappa} \equiv r'_nK^{n/2}y^n_{j,\mathrm{A,e},\kappa,i}$,

$\xi''''_{j,i,\kappa} \equiv r'_{n-1}K^{n/2}y^{n-1}_{j,\mathrm{A,e},\kappa,i}y'_{j,\mathrm{A,e},\kappa,i-1}$,

$\tilde{\xi}''_{j,i} \equiv r'r''K\tilde{y}^{(\mathrm{c})}_{j,\mathrm{A},i} + r'K\tilde{y}'^{(\mathrm{c})}_{j,\mathrm{A,c},i} + r'K\tilde{y}'^{(\mathrm{c})}_{j,\mathrm{A,e},i}$,

$$\bar{Y}_{j,\mathrm{A,e},k,i} \equiv \begin{Bmatrix} \bar{w}_{j,\mathrm{A}}K^{1/2}, & k = 1 \\ r''\bar{w}_{j,\mathrm{A}}K\bar{y}_{j,\mathrm{A,e},k-1,i+1} + \bar{w}_{j,\mathrm{A}}Ky'_{j,\mathrm{A,e},k-1,I}, & 1 < k \le N_\mathrm{A} \end{Bmatrix},$$

$$\bar{Y}_{j,\mathrm{A,c},k,i} \equiv \begin{Bmatrix} \bar{\xi}_{j,i}, & k = 0, n = r \\ \bar{\xi}_{j,i} + r''_{n-1}\bar{w}_{j,\mathrm{A}}K^{n/2}\bar{y}^{n-1}_{j,\mathrm{A,e},N_\mathrm{A},i+1}, & k = 0, n < r \\ r''\bar{w}_{j,\mathrm{A}}K\bar{y}_{j,\mathrm{A,c},k-1,i+1} + \bar{w}_{j,\mathrm{A}}Ky'_{j,\mathrm{A,c},k-1,I}, & 0 < k < N_A \end{Bmatrix},$$

where $\bar{\xi}_{j,i} \equiv r''_{n-2}\bar{w}_{j,\mathrm{A}}K^{n/2}\bar{y}^{n-2}_{j,\mathrm{A,e},N_\mathrm{A},i+1}y'_{j,\mathrm{A,e},N_\mathrm{A},I}$,

$$
\overline{X}_{j,\mathrm{A},i} \equiv \left\{ \begin{array}{l}
\overline{\xi}_{j,i+1,0}\,, \quad N_{\mathrm{A}} = 1, n = r \\
\overline{\xi}_{j,i+1,0} + \overline{\xi}'_{j,i+1,1}\,, \quad N_{\mathrm{A}} = 1, n = r' \\
\overline{\xi}_{j,i+1,0} + \overline{\xi}''_{j,i+1,1} + \overline{\xi}'''_{j,i+1,1}\,, \quad N_{\mathrm{A}} = 1, n < r' \\
\overline{\xi}_{j,i+1,N_{\mathrm{A}}-1} + \tilde{\xi}_{j,i+1}\,, \quad N_{\mathrm{A}} > 1, n = r \\
\overline{\xi}_{j,i+1,N_{\mathrm{A}}-1} + \overline{\xi}'_{j,i+1,N_{\mathrm{A}}} + \tilde{\xi}_{j,i+1}\,, \quad N_{\mathrm{A}} > 1, n = r' \\
\overline{\xi}_{j,i+1,N_{\mathrm{A}}-1} + \overline{\xi}''_{j,i+1,N_{\mathrm{A}}} + \overline{\xi}'''_{j,i+1,N_{\mathrm{A}}} + \tilde{\xi}_{j,i+1}\,, \quad N_{\mathrm{A}} > 1, n < r'
\end{array} \right\},
$$

where $\overline{\xi}_{j,i,\kappa} \equiv r''\overline{w}_{j,\mathrm{A}} K^{1/2} \overline{y}_{j,\mathrm{A,c},\kappa,i} + \overline{w}_{j,\mathrm{A}} K^{1/2} y'_{j,\mathrm{A,c},\kappa,I}$ ,

$$
\overline{\xi}'_{j,i,\kappa} \equiv \overline{w}_{j,\mathrm{A}} K^{r'/2} \overline{y}^{\,r''}_{j,\mathrm{A,e},\kappa,i} y'_{j,\mathrm{A,e},\kappa,I}\,,
$$

$$
\overline{\xi}''_{j,i,\kappa} \equiv r''_{n}\overline{w}_{j,\mathrm{A}} K^{n/2} \overline{y}^{\,n}_{j,\mathrm{A,e},\kappa,i}\,,
$$

$$
\overline{\xi}'''_{j,i,\kappa} \equiv r''_{n-1}\overline{w}_{j,\mathrm{A}} K^{n/2} \overline{y}^{\,n-1}_{j,\mathrm{A,e},\kappa,i} y'_{j,\mathrm{A,e},\kappa,I}\,,
$$

$$
\tilde{\xi}_{j,i} \equiv r''\overline{w}_{j,\mathrm{A}} K \overline{\tilde{y}}'^{(\mathrm{c})}_{j,\mathrm{A,e},i} + r''\overline{w}_{j,\mathrm{A}} K \overline{\tilde{y}}'^{(\mathrm{c})}_{j,\mathrm{A,c},i} + r''r'''\overline{w}_{j,\mathrm{A}} K \overline{\tilde{y}}^{(\mathrm{c})}_{j,\mathrm{A},i}\,,
$$

$$
Y_{j,\mathrm{A,e},k,I} \equiv \left\{ \begin{array}{l}
\overline{w}_{j,A} K^{1/2}, \quad k = 1 \\
r'\overline{w}_{j,\mathrm{A}} K y_{j,\mathrm{A,e},k-1,0}\,, \quad 1 < k \le N_{A}
\end{array} \right\},
$$

$$
Y_{j,\mathrm{A,c},k,I} \equiv \left\{ \begin{array}{l}
r'_{n-1}\overline{w}_{j,\mathrm{A}} K^{n/2} \overline{y}^{\,n-1}_{j,\mathrm{A,e},N_{\mathrm{A}},0}\,, \quad k = 0 \\
r'\overline{w}_{j,\mathrm{A}} K y_{j,\mathrm{A,c},k-1,0}\,, \quad 0 < k < N_{\mathrm{A}}
\end{array} \right\},
$$

$$
X_{j,\mathrm{A},I} \equiv \left\{ \begin{array}{l}
\psi_{j,0}\,, \quad N_{\mathrm{A}} = 1, n = r \\
\psi_{j,0} + \overline{\psi}_{j,1}\,, \quad N_{\mathrm{A}} = 1, n < r \\
\psi_{j,N_{\mathrm{A}}-1} + \tilde{\psi}_{j}\,, \quad N_{\mathrm{A}} > 1, n = r \\
\psi_{j,N_{\mathrm{A}}-1} + \overline{\psi}_{j,N_{\mathrm{A}}} + \tilde{\psi}_{j}\,, \quad N_{\mathrm{A}} > 1, n < r
\end{array} \right\},
$$

where $\psi_{j,\kappa} \equiv r'\overline{w}_{j,\mathrm{A}} K^{1/2} \overline{y}_{j,\mathrm{A,c},\kappa,0}$ ,

$$
\overline{\psi}_{j,\kappa} \equiv r'_{n}\overline{w}_{j,\mathrm{A}} K^{n/2} \overline{y}^{\,n}_{j,\mathrm{A,e},\kappa,0}\,,
$$

$$
\tilde{\psi}_{j} \equiv r'r''\overline{w}_{j,\mathrm{A}} K \overline{\tilde{y}}^{(\mathrm{c})}_{j,\mathrm{A},0}\,,
$$

$$\bar{Y}'_{j,\mathrm{A,e},k,0} \equiv \begin{Bmatrix} \bar{w}_{j,\mathrm{A}} K^{1/2}, & k=1 \\ r'' \bar{w}_{j,\mathrm{A}} K \bar{y}_{j,\mathrm{A,e},k-1,0} + \bar{w}_{j,\mathrm{A}} K y'_{j,\mathrm{A,e},k-1,I}, & 1 < k \le N_\mathrm{A} \end{Bmatrix},$$

$$\bar{Y}'_{j,\mathrm{A,c},k,0} \equiv \begin{Bmatrix} \bar{w}_{j,\mathrm{A}} K^{r/2} \bar{y}^{r''}_{j,\mathrm{A,e},N_\mathrm{A},0} y'_{j,\mathrm{A,e},N_\mathrm{A},I}, & k=0, n=r \\ r''_{n-1} \bar{w}_{j,\mathrm{A}} K^{n/2} \bar{y}^{n-1}_{j,\mathrm{A,e},N_\mathrm{A},0} + r''_{n-2} \bar{w}_{j,\mathrm{A}} K^{n/2} \bar{y}^{n-2}_{j,\mathrm{A,e},N_\mathrm{A},0} y'_{j,\mathrm{A,e},N_\mathrm{A},I}, & k=0, n<r \\ r'' \bar{w}_{j,\mathrm{A}} K \bar{y}_{j,\mathrm{A,c},k-1,0} + \bar{w}_{j,\mathrm{A}} K y'_{j,\mathrm{A,c},k-1,I}, & 0 < k < N_\mathrm{A} \end{Bmatrix},$$

$$\bar{X}'_{j,\mathrm{A},0} \equiv \begin{Bmatrix} \bar{\psi}'_{j,0}, & N_\mathrm{A} = 1, n=r \\ \bar{\psi}'_{j,0} + \bar{\xi}'_{j,0,1}, & N_\mathrm{A} = 1, n=r' \\ \bar{\psi}'_{j,0} + \bar{\xi}''_{j,0,1} + \bar{\xi}'''_{j,0,1}, & N_\mathrm{A} = 1, n<r' \\ \bar{\psi}'_{j,N_\mathrm{A}-1} + \tilde{\bar{\xi}}_{j,0}, & N_\mathrm{A} > 1, n=r \\ \bar{\psi}'_{j,N_\mathrm{A}-1} + \tilde{\bar{\xi}}_{j,0} + \bar{\xi}'_{j,0,N_\mathrm{A}}, & N_\mathrm{A} > 1, n=r' \\ \bar{\psi}'_{j,N_\mathrm{A}-1} + \bar{\xi}'''_{j,0,N_\mathrm{A}} + \bar{\xi}''_{j,0,N_\mathrm{A}} + \tilde{\bar{\xi}}_{j,0}, & N_\mathrm{A} > 1, n<r' \end{Bmatrix},$$

where $\bar{\psi}'_{j,\kappa} \equiv r'' \bar{w}_{j,\mathrm{A}} K^{1/2} \bar{y}_{j,\mathrm{A,c},\kappa,0} + \bar{w}_{j,\mathrm{A}} K^{1/2} y'_{j,\mathrm{A,c},\kappa,I}$.

Given the above definitions, the RR’s between PPF ratios can be expressed as below:

$$x_{j,\gamma,i} = X_{j,\gamma,i} / Q_{j,1,i}, \quad y_{j,\mathrm{A,e},k,i} = Y_{j,\mathrm{A,e},k,i} / Q_{j,1,i}, \quad y_{j,\mathrm{A,c},k,i} = Y_{j,\mathrm{A,c},k,i} / Q_{j,1,i}, \tag{A1}$$

$$x'_{j,\gamma,i} = X'_{j,\gamma,i} / Q'_{j,1,i}, \quad y'_{j,\mathrm{A,e},k,i} = Y'_{j,\mathrm{A,e},k,i} / Q'_{j,1,i}, \quad y'_{j,\mathrm{A,c},k,i} = Y'_{j,\mathrm{A,c},k,i} / Q'_{j,1,i}, \tag{A2}$$

$$\bar{x}_{j,\gamma,i} = \bar{X}_{j,\gamma,i} / \bar{Q}_{j,1,i}, \quad \bar{y}_{j,\mathrm{A,e},k,i} = \bar{Y}_{j,\mathrm{A,e},k,i} / \bar{Q}_{j,1,i}, \quad \bar{y}_{j,\mathrm{A,c},k,i} = \bar{Y}_{j,\mathrm{A,c},k,i} / \bar{Q}_{j,1,i}, \tag{A3}$$

$$x_{j,\gamma,I} = X_{j,\gamma,I} / Q_{j,1,I}, \quad y_{j,\mathrm{A,e},k,I} = Y_{j,\mathrm{A,e},k,I} / Q_{j,1,I}, \quad y_{j,\mathrm{A,c},k,I} = Y_{j,\mathrm{A,c},k,I} / Q_{j,1,I}, \tag{A4}$$

$$\bar{x}'_{j,\gamma,0} = \bar{X}'_{j,\gamma,0} / \bar{Q}'_{j,1,0}, \quad \bar{y}'_{j,\mathrm{A,e},k,0} = \bar{Y}'_{j,\mathrm{A,e},k,0} / \bar{Q}'_{j,1,0}, \quad \bar{y}'_{j,\mathrm{A,c},k,0} = \bar{Y}'_{j,\mathrm{A,c},k,0} / \bar{Q}'_{j,1,0}. \tag{A5}$$

We now move on to the expressions for various densities in terms of the ratios. We define below various quantities intended to simplify these expressions.

$$Q_{j,2,i} \equiv \frac{Z}{U_{j,i}U'_{j,i}}$$

$$= \sum_{\mu\nu} w_{\mu\nu} x_{j,\mu,i} x'_{j,\nu,i} + \sum_{k=1}^{N_{\mathrm{A}}-1} y_{j,\mathrm{A,e},k,i} y'_{j,\mathrm{A,c},N_{\mathrm{A}}-k,i} + \sum_{k=1}^{N_{\mathrm{A}}-1} y_{j,\mathrm{A,c},N_{\mathrm{A}}-k,i} y'_{j,\mathrm{A,e},k,i}$$

$$+ \sum_{k=1}^{N_{\mathrm{B}}-1} y_{j,\mathrm{B},k,i} y'_{j,\mathrm{B},N_{\mathrm{B}}-k,i},$$

$$z_{j,1,\mathrm{A},i} \equiv \frac{Z_{j,1,\mathrm{A},i}}{U^{r}_{j,i+1}U'_{j,i}} = K^{1/2} y'_{j,\mathrm{A,c},N_{\mathrm{A}}-1,i} + rK^{1/2} y_{j,\mathrm{A,c},N_{\mathrm{A}}-1,i+1}, \quad 0 \le i < I,$$

$$z_{j,1,\mathrm{A},I} \equiv \frac{Z_{j,1,\mathrm{A},I}}{\overline{U}^{r'}_{j,0}U'_{j,I}} = \overline{w}_{j,\mathrm{A}} K^{1/2} y'_{j,\mathrm{A,c},N_{\mathrm{A}}-1,I} + r'\overline{w}_{j,\mathrm{A}} K^{1/2} \overline{y}_{j,\mathrm{A,c},N_{\mathrm{A}}-1,0},$$

$$z_{j,2,\mathrm{A},i} \equiv \frac{Z_{j,2,\mathrm{A},i}}{U^{r}_{j,i+1}U'_{j,i}} = Kr_2 \tilde{y}^{(\mathrm{c})}_{j,\mathrm{A},i+1} + r\tilde{y}'^{(\mathrm{c})}_{j,\mathrm{A,c},i+1} + r\tilde{y}'^{(\mathrm{c})}_{j,\mathrm{A,e},i+1}, \quad 0 \le i < I,$$

$$z_{j,2,\mathrm{A},I} \equiv \frac{Z_{j,2,\mathrm{A},I}}{\overline{U}^{r'}_{j,0}U'_{j,I}} = Kr'_2 \overline{w}_{j,\mathrm{A}} \overline{\tilde{y}}^{(\mathrm{c})}_{j,\mathrm{A},0} + r'\overline{w}_{j,\mathrm{A}} \overline{\tilde{y}}'^{(\mathrm{c})}_{j,\mathrm{A,c},0} + r'\overline{w}_{j,\mathrm{A}} \overline{\tilde{y}}'^{(\mathrm{c})}_{j,\mathrm{A,e},0},$$

$$z_{j,c,i} \equiv \frac{Z_{j,c,i}}{U^{r}_{j,i+1}U'_{j,i}} = r_{n-1} K^{n/2} y^{n-1}_{j,\mathrm{A,e},N_{\mathrm{A}},i+1} y'_{j,\mathrm{A,e},N_{\mathrm{A}},i} + r_n K^{n/2} y^{n}_{j,\mathrm{A,e},N_{\mathrm{A}},i+1}, \quad 0 \le i < I,$$

$$z_{j,c,I} \equiv \frac{Z_{j,c,I}}{\overline{U}^{r'}_{j,0}U'_{j,I}} = \left\{ \begin{array}{l} K^{r/2} \overline{y}^{r'}_{j,\mathrm{A,e},N_{\mathrm{A}},0} y'_{j,\mathrm{A,e},N_{\mathrm{A}},I}, \quad n = r \\ r'_{n-1} K^{n/2} \overline{y}^{n-1}_{j,\mathrm{A,e},N_{\mathrm{A}},0} y'_{j,\mathrm{A,e},N_{\mathrm{A}},I} \\ \quad + r'_n K^{n/2} \overline{y}^{n}_{j,\mathrm{A,e},N_{\mathrm{A}},0}, \quad n < r \end{array} \right\}.$$

In terms of the above quantities, the endpoint, bifunctionality, and core densities can be expressed as

$$\phi_{j,1,\mathrm{A},i} = \frac{z_{j,1,\mathrm{A},i}}{Q_{j,1,i}Q_{j,2,i}}, \quad \phi_{j,2,\mathrm{A},i} = \frac{z_{j,2,\mathrm{A},i}}{Q_{j,1,i}Q_{j,2,i}}, \quad \phi_{j,c,i} = \frac{z_{j,c,i}}{Q_{j,1,i}Q_{j,2,i}}.$$

TABLE I. Surface excesses of stars ($\sigma_A$), linear chains ($\sigma_B$), and free volume ($\sigma_C$) for 3-arm star-linear blends with the designated bulk free volume densities ($\phi_{C,\infty}$)

| $\phi_{C,\infty}$ | $\sigma_A$ | $\sigma_B$ | $\sigma_C$ |
|---|---|---|---|
| 0 | 0.254 | -0.254 | 0 |
| 0.30 | 0.482 | -0.511 | 0.029 |
| 0.50 | 0.781 | -0.877 | 0.096 |

TABLE II. Surface excesses of stars ($\sigma_A$), linear chains ($\sigma_B$), and free volume ($\sigma_C$) for 4-arm star-linear blends with the designated bulk free volume densities ($\phi_{C,\infty}$)

| $\phi_{C,\infty}$ | $\sigma_A$ | $\sigma_B$ | $\sigma_C$ |
|---|---|---|---|
| 0 | -0.136 | 0.136 | 0 |
| 0.30 | 0.488 | -0.510 | 0.022 |
| 0.50 | 1.223 | -1.302 | 0.079 |

TABLE III. Surface excesses of stars ($\sigma_A$), linear chains ($\sigma_B$), and free volume ($\sigma_C$) for 5-arm star-linear blends with the designated bulk free volume densities ($\phi_{C,\infty}$)

| $\phi_{C,\infty}$ | $\sigma_A$ | $\sigma_B$ | $\sigma_C$ |
|---|---|---|---|
| 0 | -1.493 | 1.493 | 0 |
| 0.30 | -0.117 | 0.104 | 0.013 |
| 0.50 | 1.245 | -1.305 | 0.061 |

LIST OF FIGURES AND CAPTIONS

FIGURE 1: A schematic diagram of the recursive tree structure for a blend confined between two different surfaces, showing the two halves $\Im_1$ and $\Im_2$ of the bulk tree $\Im$ and their corresponding surface trees, $\overline{\Im}_1$ and $\overline{\Im}_2$.

FIGURE 2: Bulk-normalized monomer density of star plotted against distance from the surface in units of the end-to-end distance of a star arm, for athermal 3-arm star-linear blends with 199 monomers per polymer and three different bulk densities of free volume. The solid, dotted, and dashed lines represent bulk free volume densities of 0%, 30%, and 50%, respectively. The density of star is normalized by dividing out its value in the bulk. The inset shows a few levels next to the surface.

FIGURE 3: Bulk-normalized monomer density of star plotted against distance from the surface in units of the end-to-end distance of a star arm, for athermal 4-arm star-linear blends with 201 monomers per polymer and three different bulk densities of free volume. The solid, dotted, and dashed lines represent bulk free volume densities of 0%, 30%, and 50%, respectively. The density of star is normalized by dividing out its value in the bulk. The inset shows a few levels next to the surface.

FIGURE 4: Bulk-normalized monomer density of star plotted against distance from the surface in units of the end-to-end distance of a star arm, for athermal 5-arm star-linear blends with 201 monomers per polymer and three different bulk densities of free volume. The solid, dotted, and dashed lines represent bulk free volume densities of 0%, 30%, and 50%, respectively. The density of star is normalized by dividing out its value in the bulk. The inset shows a few levels next to the surface.

FIGURE 5: Bulk-normalized density of free volume (C) plotted against distance from the surface in units of the end-to-end distance of a star arm, for athermal 3-arm star-linear blends with 199 monomers per polymer and two different bulk densities of free volume. The solid and dotted lines represent bulk free volume densities of 30% and 50%, respectively. The density of C is normalized by dividing out its value in the bulk.

FIGURE 6: Bulk-normalized density of free volume (C) plotted against distance from the surface in units of the end-to-end distance of a star arm, for athermal 4-arm star-linear blends with 201 monomers per polymer and two different bulk densities of free volume. The solid and dotted lines represent bulk free volume densities of 30% and 50%, respectively. The density of C is normalized by dividing out its value in the bulk.

FIGURE 7: Bulk-normalized density of free volume (C) plotted against distance from the surface in units of the end-to-end distance of a star arm, for athermal 5-arm star-linear blends with two different bulk densities of free volume. The solid and dotted lines represent bulk free volume densities of 30% and 50%, respectively. The density of C is normalized by dividing out its value in the bulk.

FIGURE 8: Entropy density of athermal, incompressible blends of A stars (with various numbers of arms) and B chains, confined between two identical surfaces, plotted against the interaction energy between A and both surfaces. Filled circles, empty circles, filled triangles, and empty triangles represent blends of equal molecular weight B chains with 3-arm, 199-monomer A stars; 4-arm, 201-monomer A stars; 5-arm, 201-monomer A stars; and 2-arm, 201-monomer A stars, respectively.

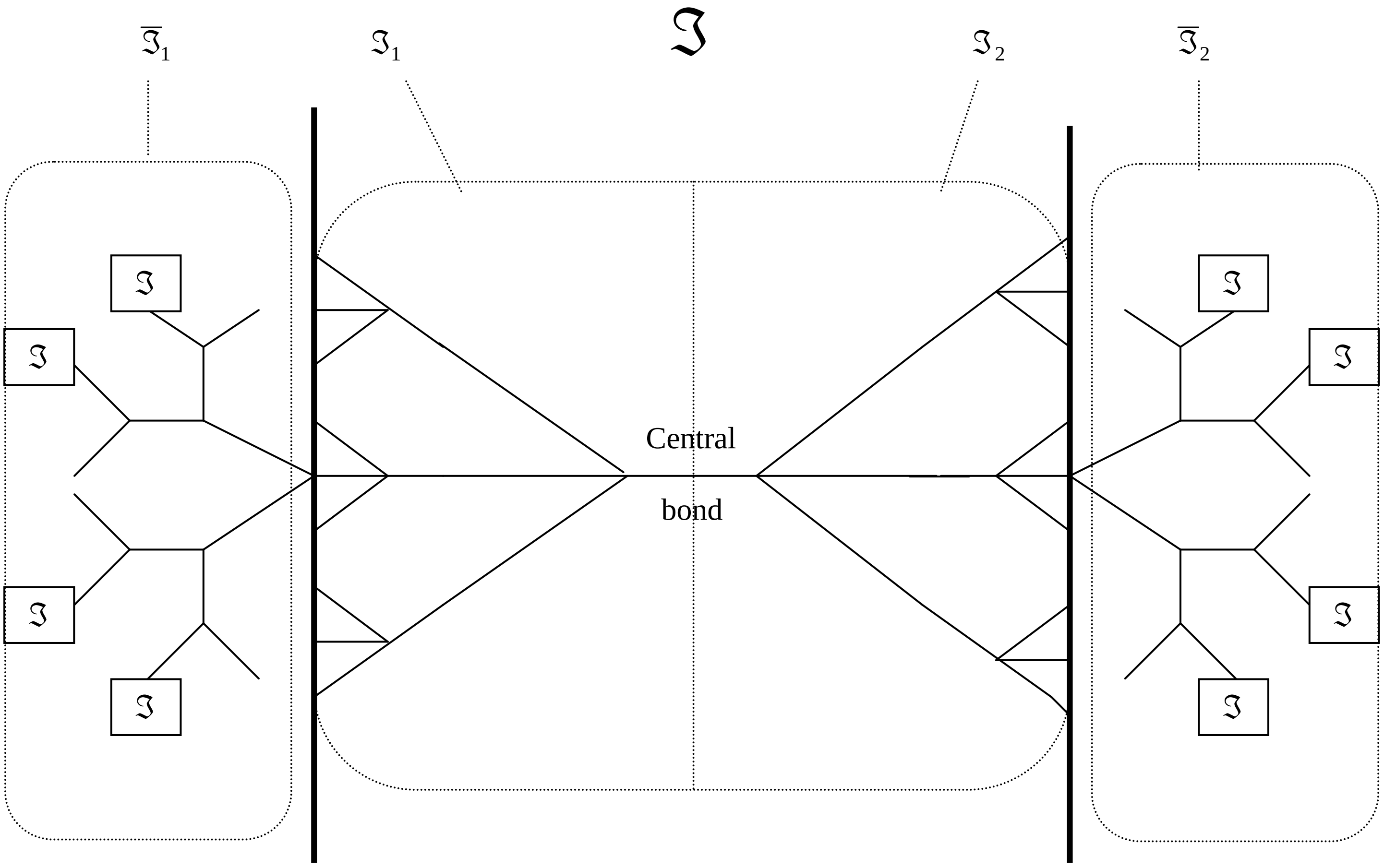

Richard Batman and P.D. Gujrati, Fig.1

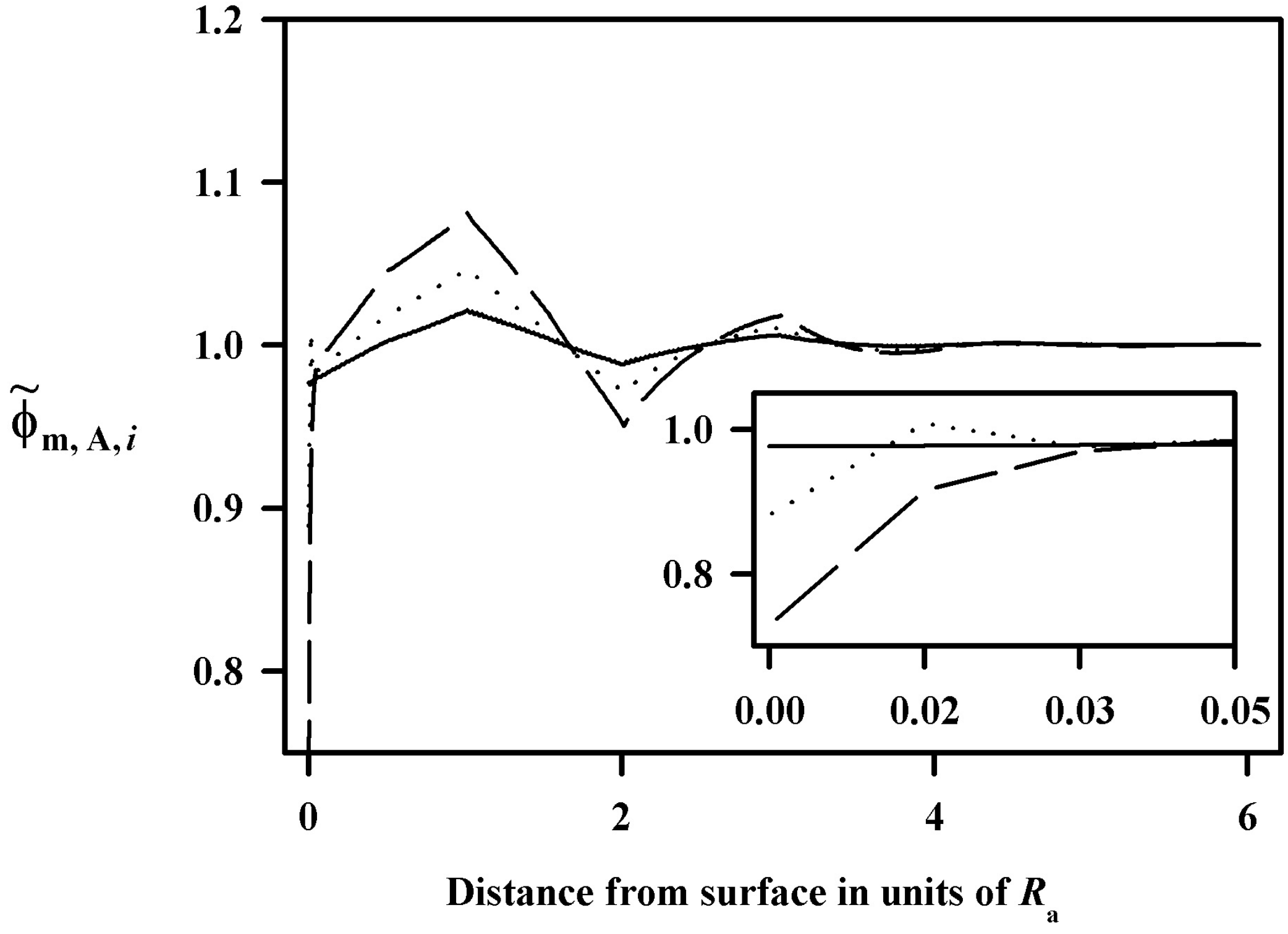

Richard Batman and P.D. Gujrati, Fig.2

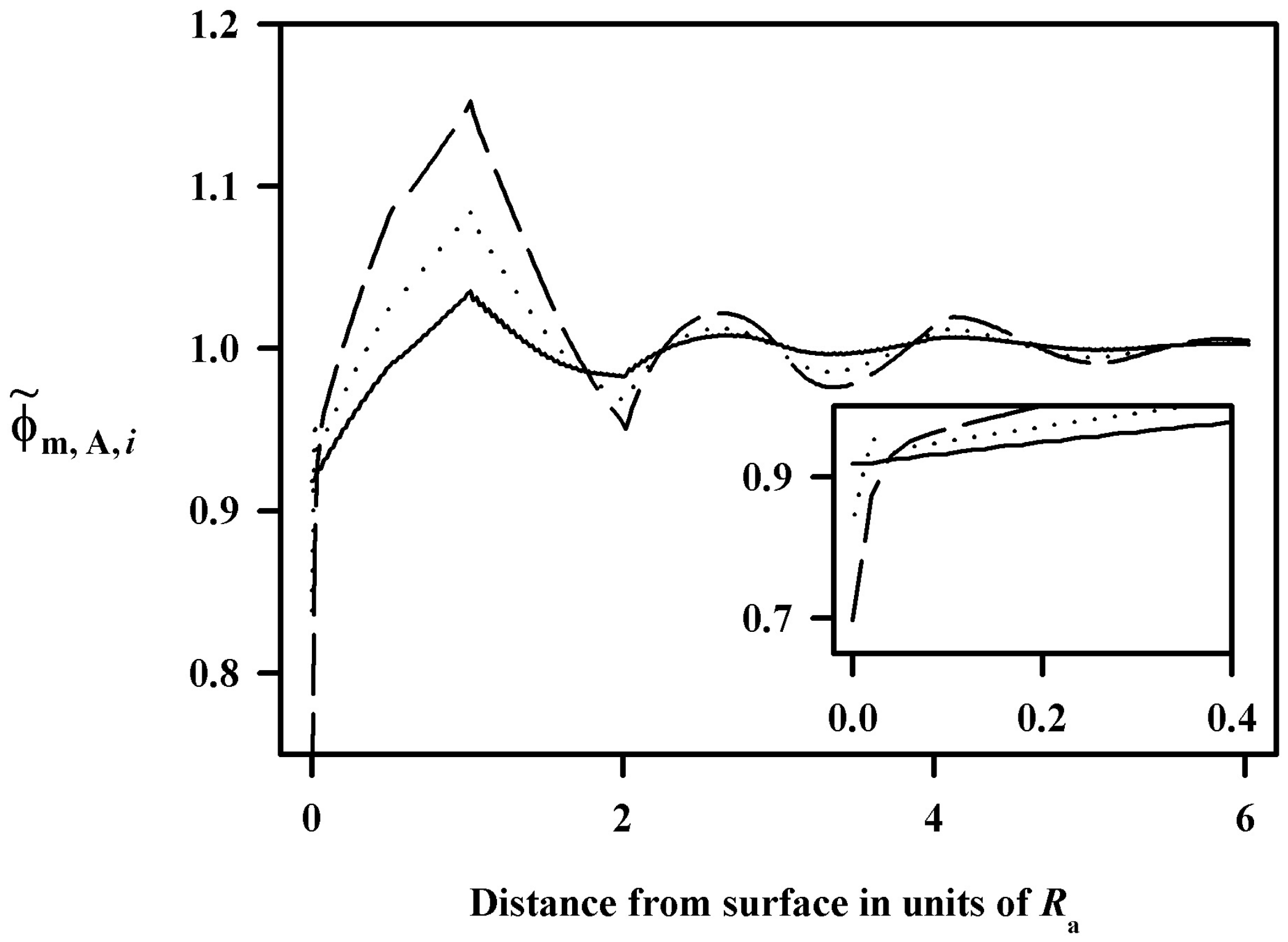

Richard Batman and P.D. Gujrati, Fig.3

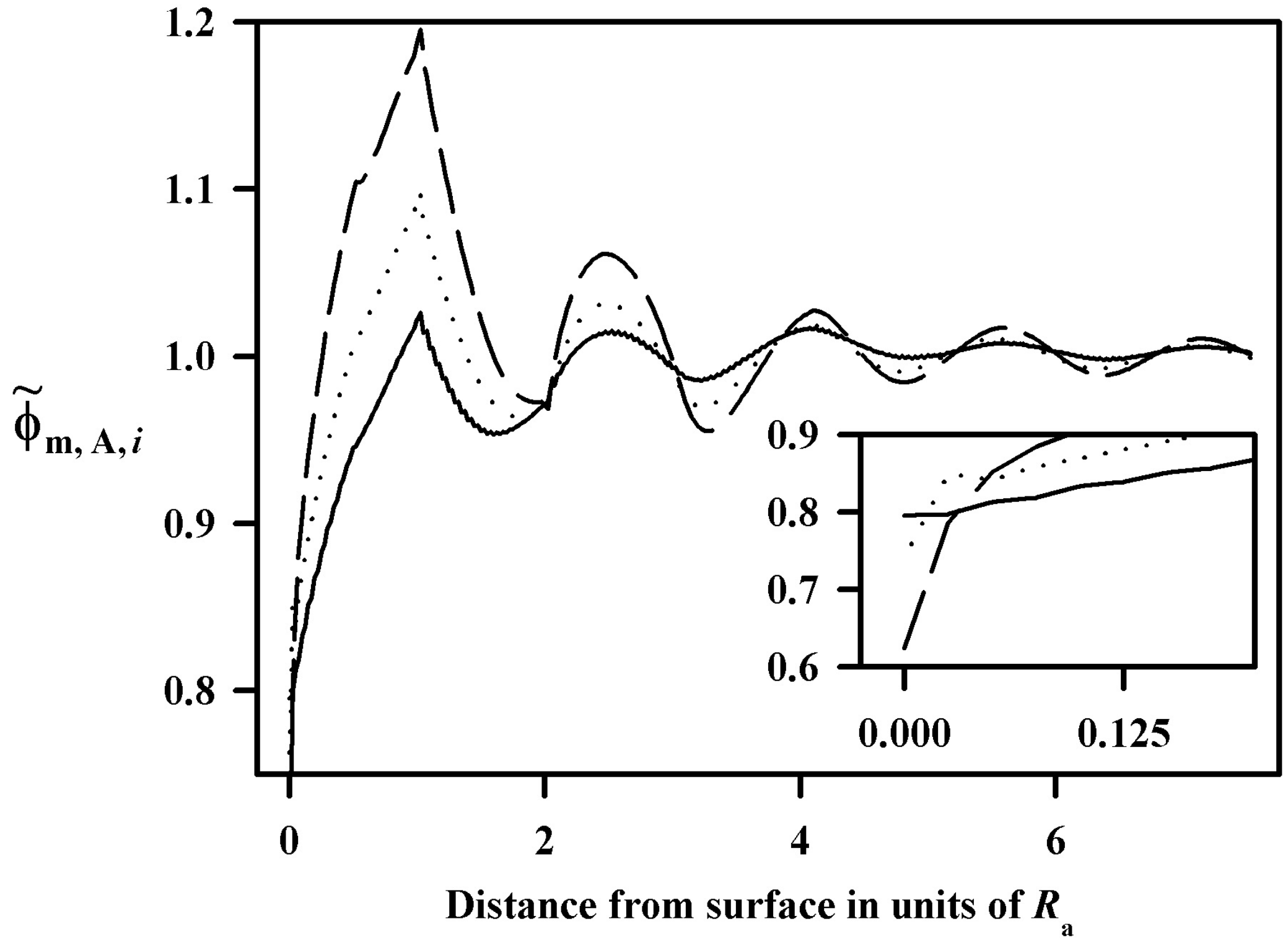

Richard Batman and P.D. Gujrati, Fig.4

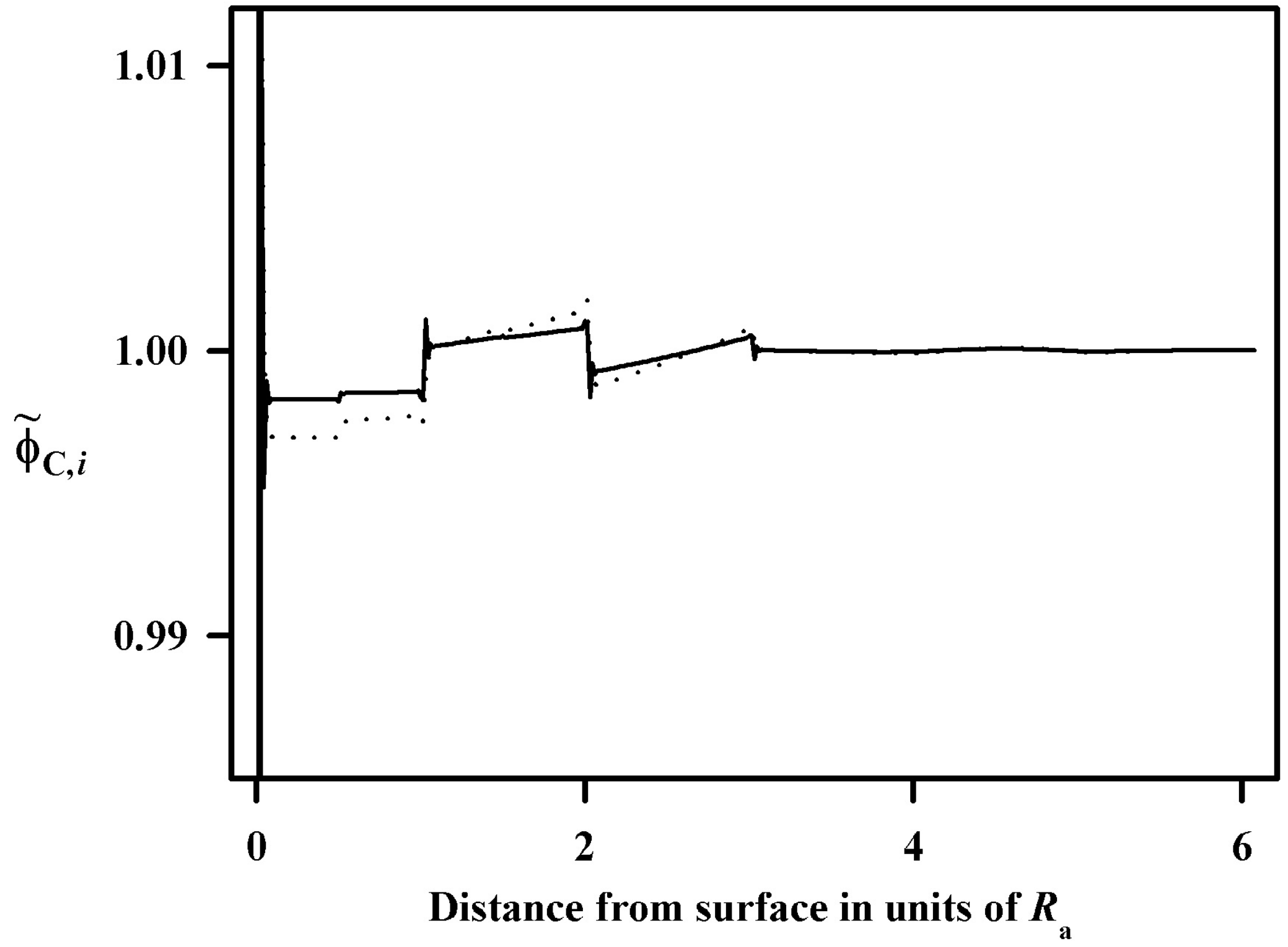

Richard Batman and P.D. Gujrati, Fig.5

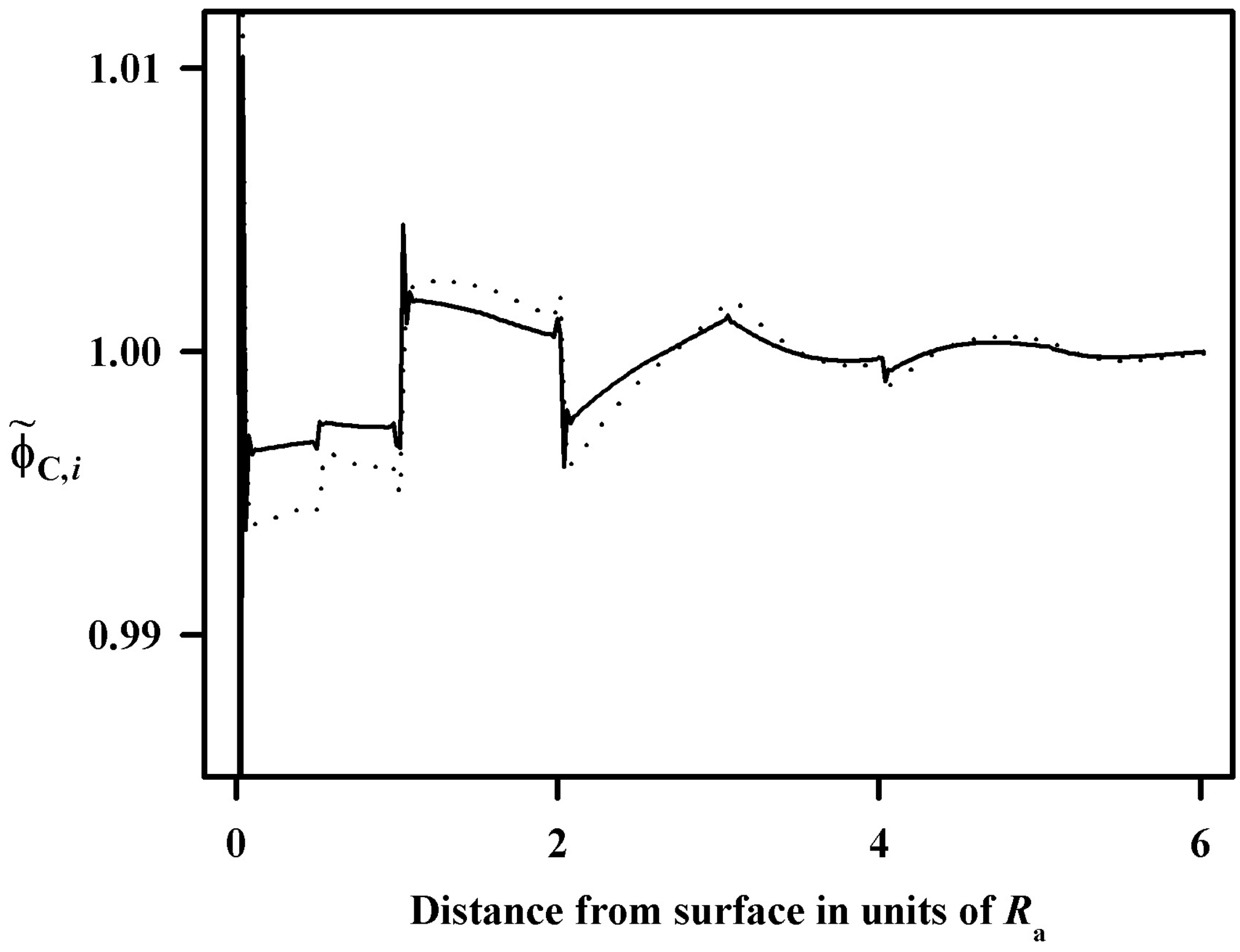

Richard Batman and P.D. Gujrati, Fig.6

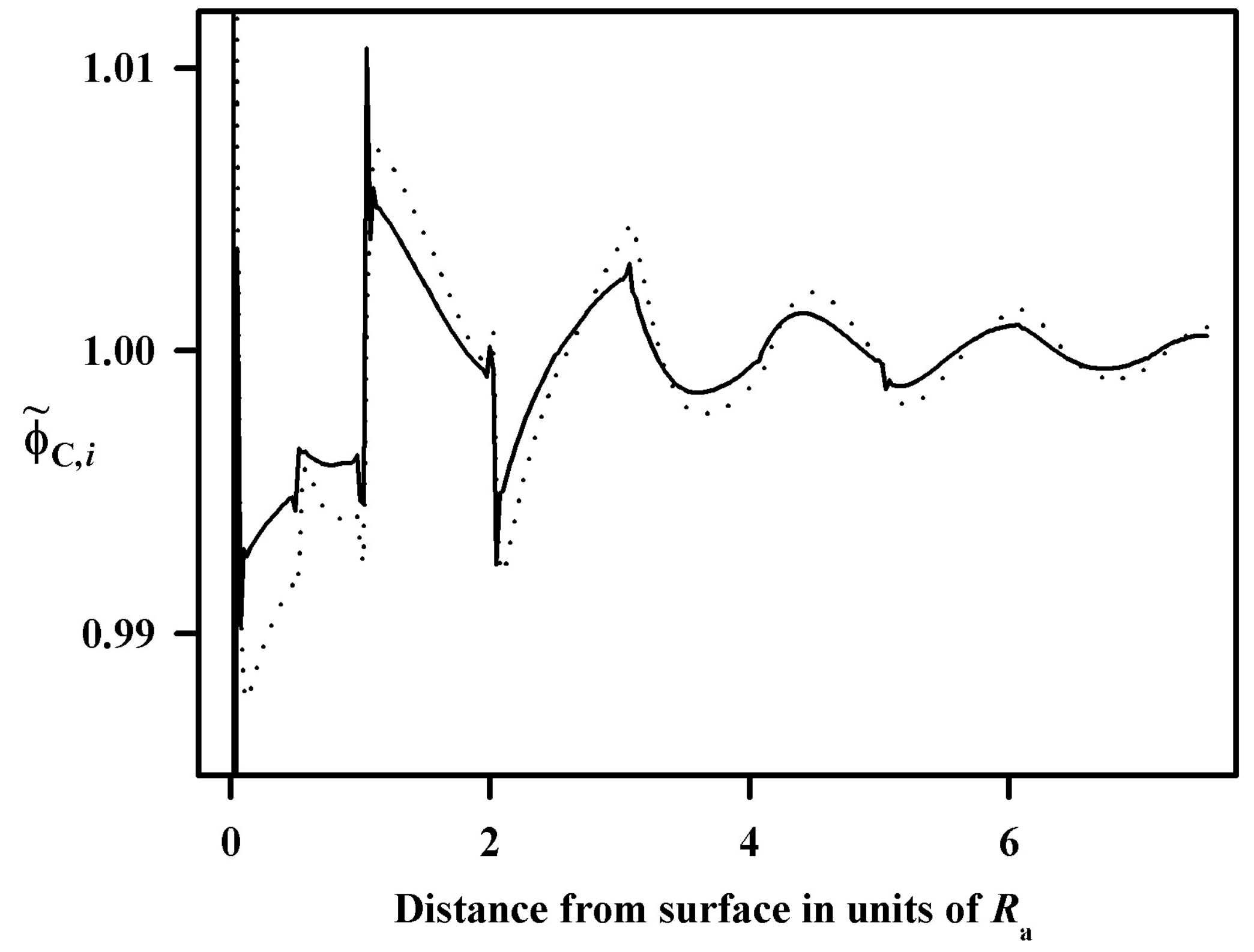

Richard Batman and P.D. Gujrati, Fig.7

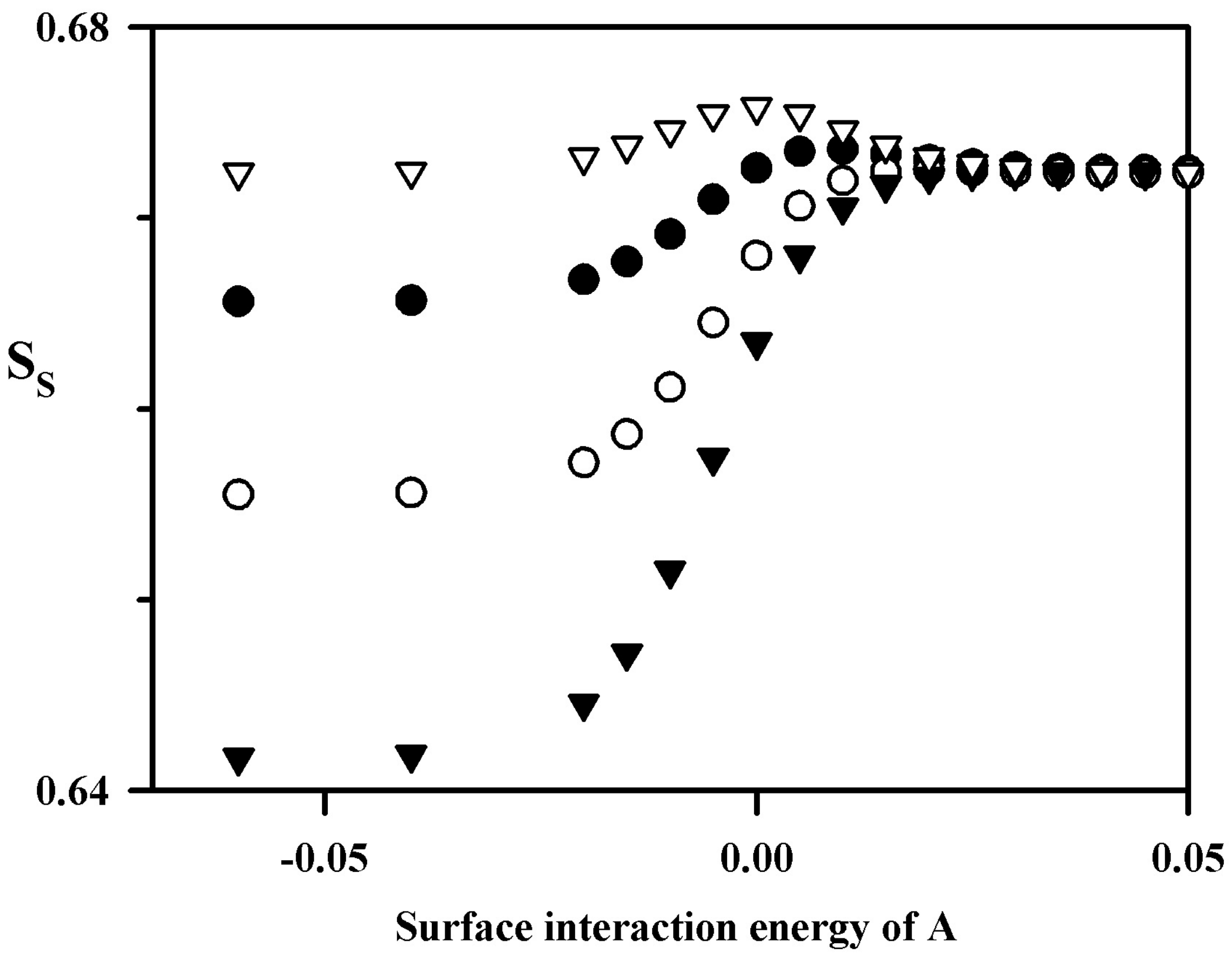

Richard Batman and P.D. Gujrati, Fig.8